\documentclass{article}

\usepackage[numbers,compress]{natbib}
\usepackage[preprint]{neurips_data_2021}

\usepackage[utf8]{inputenc} % allow utf-8 input
\usepackage[T1]{fontenc}    % use 8-bit T1 fonts
\usepackage{hyperref}       % hyperlinks
\usepackage{url}            % simple URL typesetting
\usepackage{booktabs}       % professional-quality tables
\usepackage{amsfonts}       % blackboard math symbols
\usepackage{nicefrac}       % compact symbols for 1/2, etc.
\usepackage{microtype}      % microtypography
\usepackage{xcolor}         % colors

\usepackage{amsfonts}       % blackboard math symbols
\usepackage{microtype}
\usepackage{graphicx}
\usepackage{subfigure}
\usepackage{amsmath}
\usepackage{amssymb}
\usepackage{amsthm}
\usepackage{dsfont}
\usepackage{multirow}
\usepackage{bbding}
\usepackage{caption}
\usepackage{tabularx}
\usepackage{pifont}
\usepackage{color}
\usepackage{arydshln}
\usepackage[ruled, vlined]{algorithm2e}
\usepackage[frozencache=true,cachedir=minted]{minted}
\usepackage{ulem}
\usepackage{enumitem}
\newcommand{\xb}{\ensuremath{\mathbf{x}}}
\newcommand{\rb}{\ensuremath{\mathbf{r}}}
\newcommand{\yb}{\ensuremath{\mathbf{y}}}
\newcommand{\dsname}{\ensuremath{\textsc{NaturalProofs}}}

\newcommand{\cmark}{\ding{51}}%
\newcommand{\xmark}{\ding{55}}%

\newcommand{\eos}{\ensuremath{\left<\text{eos}\right>}}

\newcommand*{\horzbar}{\rule[.5ex]{2.5ex}{0.5pt}}
\newcommand{\myparagraph}[1]{\par\noindent\textbf{{#1}.}} 

\title{\textsc{NaturalProofs}: Mathematical Theorem Proving in Natural Language}

\author{
  Sean Welleck$^{1,2}$,
  Jiacheng Liu$^{1}$,
  Ronan Le Bras$^{2}$, \\
  \textbf{Hannaneh Hajishirzi}$^{1,2}$,
  \textbf{Yejin Choi}$^{1,2}$,
  \textbf{Kyunghyun Cho}$^{3,4}$ \\
  $^1$Paul G. Allen School of Computer Science \& Engineering, University of Washington \\
  $^2$Allen Institute for Artificial Intelligence \\
  $^3$New York University \\
  $^4$CIFAR Fellow in Learning in Machines \& Brains \\
  \texttt{wellecks@uw.edu}
}

\begin{document}

\maketitle

\begin{abstract}
Understanding and creating mathematics using natural mathematical language -- the mixture of symbolic and natural language used by humans -- is a challenging and important problem for driving progress in machine learning.
As a step in this direction, we develop \textsc{NaturalProofs}, a multi-domain corpus of mathematical statements and their proofs, written in natural mathematical language.
\textsc{NaturalProofs} unifies broad coverage, deep coverage, and low-resource mathematical sources, allowing for evaluating both in-distribution and zero-shot generalization.
Using \textsc{NaturalProofs}, we benchmark strong neural methods on mathematical reference retrieval and generation tasks which test a system's ability to determine key results that appear in a proof.
Large-scale sequence models show promise 
compared to classical information retrieval methods, yet their performance and out-of-domain generalization leave substantial room for improvement.
\textsc{NaturalProofs} opens many avenues for research on challenging mathematical tasks.\footnote{Dataset and code available at \url{https://github.com/wellecks/naturalproofs}.}
\end{abstract}

\section{Introduction}
\label{sec:introduction}

Solving the problem of understanding and creating mathematics using \textit{natural mathematical language} -- the mixture of symbolic and natural language used by humans -- is a path towards developing agents capable of reasoning.
The mixture of symbolic and natural text, 
along with the existence of a formal counterpart, offers a unique setting for studying reasoning that complements research involving natural language alone or purely within a formal system.
Constructing a mathematical proof involves symbolic manipulation, logical and analogical reasoning, as well as knowledge retrieval.
Common sense and natural language abilities are needed to articulate the proof in a concise, comprehensible form.
Moreover, systems that operate on mathematical text have applications in education and scientific discovery, while bridging informal and formal mathematics can be a key driver of progress in automated reasoning \citep{carter2013lurch,kang_document-level_2020,szegedy2020promising}.

Recently, techniques from natural language processing have driven advances in \textit{formalized mathematics} (e.g. \citet{polu2020generative,rabe2021mathematical,wu2021lime}), in which mathematics is written in a verifiable formal language that resembles source code, such as  Mizar \citep{urban2006mptp},  Lean \citep{demoura2015lean}, or Metamath \citep{megill2019metamath}.
However, this setting does not directly address the \textit{informal} aspect of human mathematics, which is conveyed with a mixture of symbolic and natural language \citep{gowers2008princetoncompanion}.
This aspect is crucial, since advancing \textit{human understanding} is a goal of mathematics \citep{thurston1994proof}, and a significant fraction of mathematical knowledge is in natural language text \citep{szegedy2020promising}.

\begin{table}[t]
\footnotesize
\setlength{\tabcolsep}{2pt}
\begin{center}
\begin{tabular}{l p{12cm}}
\toprule
\textbf{Source} & \textbf{ProofWiki} \\
\hline
\textbf{Theorem} & \textbf{Category of Monoids is Category} \\
& Let $\mathrm{Mon}$ be the category of monoids. \\
& Then $\mathrm{Mon}$ is a metacategory. \\
\hline
\textbf{Proof} & Let us verify the axioms $(C1)$ up to $(C3)$ for a {\uline{metacategory}}. We have \\
& {\uline{Composite of Homomorphisms on Algebraic Structure is Homomorphism}}, verifying $(C1)$. \\
& We have {\uline{monoid}} $\left({S, \circ}\right)$. Now, $(C2)$ follows from \\
& {\uline{Identity Mapping is Left Identity and Identity Mapping is Right Identity}}. \\
& Finally, $(C3)$ follows from {\uline{Composition of Mappings is Associative}}. \\
& Hence $\mathrm{Mon}$ is a {\uline{metacategory}}. \\
\bottomrule
\addlinespace[0.2em]
\toprule
\textbf{Source} & \textbf{Textbook: Real Analysis} \\
\hline
\textbf{Theorem} & Suppose that  $f$ is continuous on the closed interval $[a,b]$ and differentiable on the \\
& open interval $(a,b),$ and $f(a)=f(b).$\\ 
& Then $f'(c)=0$ for some $c$ in the open interval $(a,b).$\\
\hline
\textbf{Proof} & Since $f$ is continuous on $[a,b]$, $f$ attains a maximum and a minimum value on $[a,b]$ (\uline{Theorem 2.2.9}). If these two extreme values are the same, then $f$ is constant on $(a,b)$, so  $f'(x)=0$ for all $x$ in $(a,b)$. If the extreme values differ, then at least one must be attained at some point $c$ in the open interval  $(a,b)$, and $f'(c)=0$, by \uline{Theorem 2.3.7}.\\
\bottomrule
\addlinespace[0.2em] 
\end{tabular}
\end{center}
\caption{
    Example theorems and their proofs from $\textsc{NaturalProofs}$. 
    Given a theorem, the mathematical retrieval task consists of retrieving the \uline{references} (underlined) that occur in its proof.
    $\textsc{NaturalProofs}$ contains data from ProofWiki, Stacks, and two textbooks; we show two sources here and two other sources in \autoref{tbl:dataset-example-more}.
    See \autoref{json-example} and \autoref{fig:schema} for data format details.
}
\label{tbl:dataset-example}
\end{table}

In this paper, we describe $\dsname$, a multi-domain corpus of mathematical statements and their proofs, written in natural mathematical language.
$\dsname$ contains \textit{broad-coverage} data from ProofWiki,\footnote{\url{https://proofwiki.org/}} \textit{deep-coverage} data from the Stacks project,\footnote{\url{https://stacks.math.columbia.edu/}} and \textit{low-resource, real-world} data from mathematics textbooks.
$\dsname$ unifies these sources in a common schema and is made publicly available as a resource to drive progress on tasks involving informal mathematics, complementing existing work in this direction (e.g. \cite{ferreira2020natural,ferreira2020premise,wang2020exploration}).

Using $\dsname$, we consider \textit{mathematical reference retrieval}, an analogue of premise selection~\citep{alemi2016deepmath,ferreira2020premise}: given a mathematical claim, retrieve the set of references (theorems, lemmas, definitions) that occur in its proof.
This task represents a crucial facet of mathematical reasoning, in which a mathematician determines the key results that appear in a proof.
As a bridge towards generative tasks using $\dsname$, we consider \textit{mathematical reference generation}, which requires additionally recovering the order and number of references in each proof.

In addition to standard \textit{in-distribution} evaluation, the multi-domain nature of $\dsname$ allows for evaluating \textit{out-of-distribution}, zero-shot generalization. 
We design an evaluation protocol that tests a system's ability to retrieve references for \textit{novel} theorems in each setting, and benchmark methods based on large-scale neural sequence models \citep{devlin2019bert,karpukhin2020dense}, including a strong \textit{joint retrieval} method that better refines the top of the ranked list, 
as well as an \textit{autoregressive} variant for reference generation.
The neural methods are effective for in-domain retrieval compared to classical techniques,
yet out-of-distribution generalization, 
leveraging symbolic mathematical content, and fully recovering a proof's references remain as fundamental challenges.
\textsc{NaturalProofs} opens many possibilities for developing and evaluating machine learning methods on challenging mathematical tasks.

\section{Related Work}
\label{sec:related-work}

\myparagraph{Machine learning for mathematical theorem proving}
A large portion of work integrating machine learning with mathematical reasoning has focused on formalized mathematics.
Early work by \citet{urban2006mptp} used machine learning for selecting relevant premises in the Mizar mathematical library that are passed to an automated theorem prover,
which was later explored with deep neural networks \citep{alemi2016deepmath}.
\citet{bansal2019holist} developed the HOList benchmark based on the HOL Light theorem prover, while other benchmark tasks use the Coq \citep{huang2019gamepad,yang2019learning}, Metamath \citep{whalen2016holophrasm,wang2020learning,polu2020generative}, or Isabelle \citep{li2021isarstep} environments.
These formalized settings differ from \textsc{NaturalProofs}, which uses mathematical language as humans write it.
\citet{szegedy2020promising} argues for leveraging both informal and formal mathematics through autoformalization.
\citet{wang2020exploration} explore translating between informal and formal mathematics, including via a dataset based on ProofWiki, though their dataset is not made available.
\citet{ferreira2020natural,ferreira2020premise} propose a classification-based natural language premise selection task and a dataset based on ProofWiki, while $\dsname$ covers multiple domains and provides evaluation and benchmarks for full retrieval and generative tasks.

\myparagraph{Mathematics and language benchmarks}
Several datasets evaluate a model's ability to solve multiple-choice algebraic word problems \citep{roy2015solving,ling2017program,amini2019mathqa} or arithmetic problems \citep{saxton2018analysing} with varying degrees of natural language. 
\citet{lample2020deep} evaluate neural sequence models on symbolic integration problems, while \citet{hendrycks2021measuring} propose a benchmark based on math competition problems.
\textsc{NaturalProofs} focuses on theorem proving rather than calculation, which we hypothesize evaluates different skills, and may prove useful in bridging formal and informal settings.

\myparagraph{Large-scale neural language models}
Large-scale unsupervised pretraining of language models has led to significant advances in many natural language processing domains (e.g. \cite{devlin2019bert,radford2019language,raffel2020t5,brown2020gpt3}).
Recent work suggests that these models store knowledge in their parameters \citep{petroni2020language}, are capable of reasoning in mathematical \citep{rabe2021mathematical,wu2021lime} and language \citep{clark2020transformers,tafjord2020proofwriter} domains, and are effective for information retrieval tasks \citep{nogueira2020passage,nogueira2020beyond}.
These advances motivate our work, which explores mathematical reasoning in natural language with large-scale language models through a retrieval task.

\section{The \textsc{NaturalProofs} Dataset}
\label{sec:dataset}

\begin{table}[t]
\begin{minipage}{.48\linewidth}
\centering
\includegraphics[width=\columnwidth]{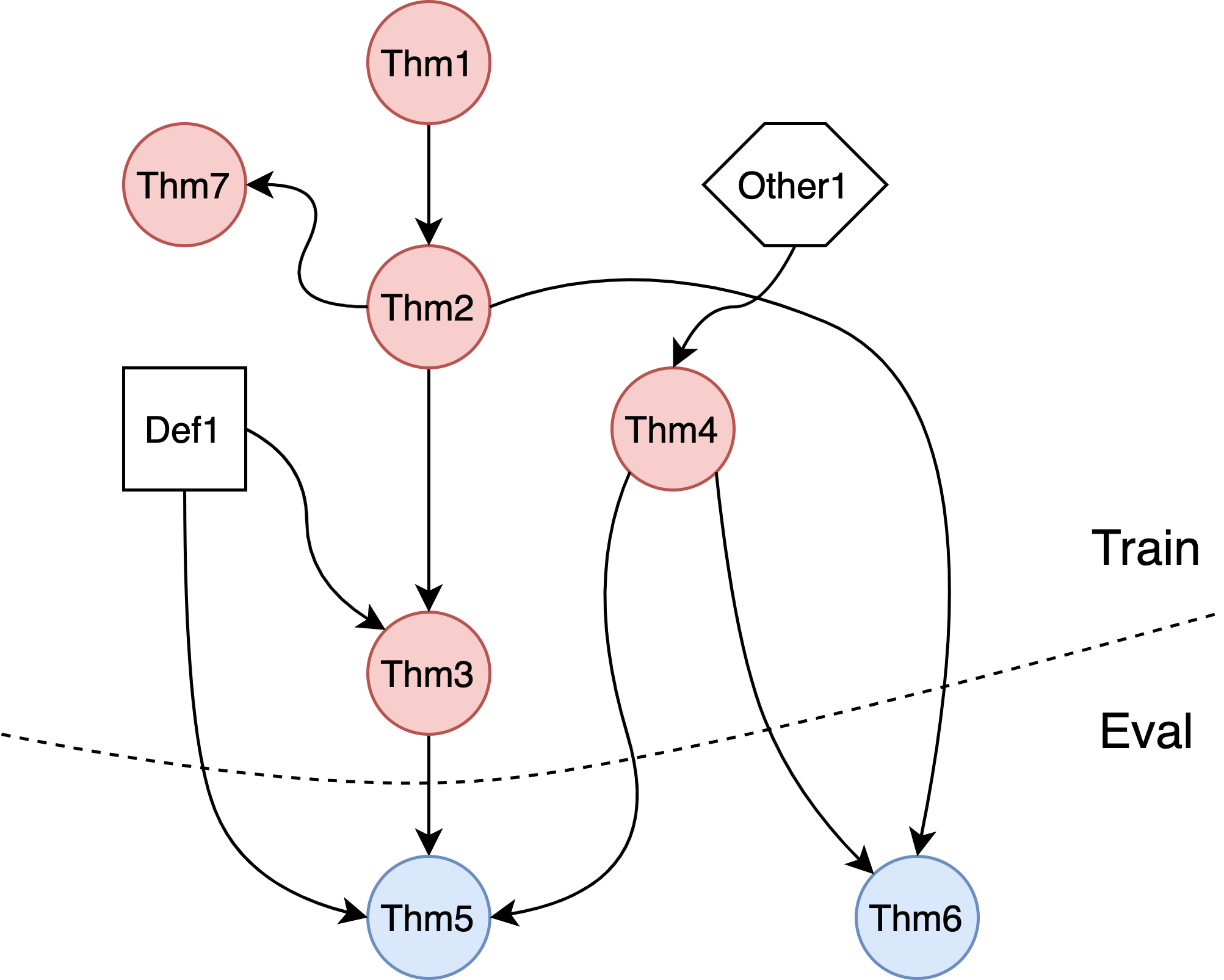}
\vspace{5pt}
\caption{
    The reference graph. 
    Nodes are \textit{statements} and edges are \textit{reference} links.
    An edge pointing from A to B means that the proof for \textit{theorem} B refers to \textit{statement} A.
    Edges can start from any type of \textit{statement}, but they always end at a \textit{theorem}.
    In our tasks, the dataset is split so that all theorems in the evaluation sets are \textit{leaf} nodes in the reference graph. 
}
\label{fig:ref-graph}
\end{minipage}
\hfill
\begin{minipage}{.48\linewidth}
\setlength{\tabcolsep}{3pt}
\begin{center}
\footnotesize
\begin{tabular}{rr|r|rrrr}
\toprule
 & \textbf{Source} & \textbf{All} & \textbf{PWiki} & \textbf{Stacks} & \textbf{RA} & \textbf{NT} \\
\toprule
\multirow{4}{*}{\rotatebox[origin=c]{90}{\textbf{Theorem}}}
                     & N       & \textit{32,579} & \textit{19,734} & \textit{12,479} & \textit{298} & \textit{68} \\
                     & Tokens  & 46.7 & 38.2 & 60.6 & 33.6 & 23.7 \\
                     & Lines   & 5.9 & 3.6 & 9.7 & 8.4 & 4.5 \\
                     & Refs    & 1.8 & 2.8 & 0.2 & 0.0 & 0.0 \\
                     \midrule
\multirow{4}{*}{\rotatebox[origin=c]{90}{\textbf{Proof}}}
                     & N       & \textit{32,012} & \textit{19,234} & \textit{12,479} & \textit{235} & \textit{64} \\
                     & Tokens  & 181.5 & 199.3 & 155.5 & 128.9 & 97.2 \\
                     & Lines   & 24.9 & 25.8 & 23.4 & 36.1 & 16.1 \\
                     & Refs    & 5.6 & 7.4 & 3.0 & 1.6 & 0.9 \\
                     \midrule
\multirow{4}{*}{\rotatebox[origin=c]{90}{\textbf{Definition}}}
                     & N       & \textit{14,230} & \textit{12,420} & \textit{1,687} & \textit{86} & \textit{37} \\
                     & Tokens  & 48.4 & 45.0 & 73.2 & 58.6 & 32.6 \\
                     & Lines   & 5.0 & 4.2 & 10.7 & 13.3 & 5.1 \\
                     & Refs    & 2.9 & 3.3 & 0.4 & 0.0 & 0.0 \\
                     \midrule
\multirow{4}{*}{\rotatebox[origin=c]{90}{\textbf{Other}}}
                     & N       & \textit{1,974} & \textit{1,006} & \textit{968} & -- & -- \\
                     & Tokens  & 212.1 & 286.1 & 135.2 & -- & -- \\
                     & Lines   & 34.4 & 46.7 & 21.7 & -- & -- \\
                     & Refs    & 5.7 & 9.2 & 2.0 & -- & -- \\
\bottomrule
\end{tabular}
\end{center}
\caption{
    \textsc{NaturalProofs} dataset statistics. Numbers represent mean value, except for "N" rows which represent count.
    \textbf{RA} is the Real Analysis textbook; \textbf{NT} is the Number Theory textbook. 
    See \autoref{tbl:dataset-stats-detail} for detailed statistics.
}
\label{tbl:dataset-stats}
\end{minipage}
\end{table}

The \textsc{NaturalProofs} Dataset is a large-scale, multi-domain dataset for studying mathematical reasoning in natural language.
\textsc{NaturalProofs} consists of 32k theorem statements and proofs, 14k definitions, and 2k other types of pages (e.g. axioms, corollaries)
derived from three domains: \textit{broad-coverage} data from ProofWiki, an online compendium of mathematical proofs written by a community of contributors; \textit{deep-coverage} data from the Stacks project, a collaborative web-based textbook of algebraic geometry; and \textit{low-resource, real-world} data from mathematics textbooks.
\autoref{tbl:dataset-example} shows example theorems and proofs from \textsc{NaturalProofs}, and \autoref{tbl:dataset-stats} shows statistics.

\myparagraph{Multi-domain} \textsc{NaturalProofs} provides a common schema for mathematical statements, proofs, and the references that appear in each.
Its multiple domains provide a challenging evaluation setting for models and opens opportunities for investigating domain transfer, out-of-distribution generalization, and methods for low-resource settings.
This differs from existing resources that focus only on ProofWiki \citep{ferreira2020natural,ferreira2020premise}, and reflects shifts in natural language processing towards multi-domain settings \citep{williams2018mnli,hu2020xtreme}, out-of-distribution generalization \citep{lebras2020adversarial,hendrycks2020pretrained,thakur2021beir}, 
and few- or zero-shot generalization in resource-constrained settings \citep{brown2020gpt3,ebrahimi2021americasnli}.

\textbf{Structure.}
Each \textit{statement} in \textsc{NaturalProofs} 
is either a theorem or a definition.
\textsc{NaturalProofs} provides the statement's title, contents, and references.
The \textit{contents} is a list of sequences, where each sequence contains one line of mixed text and \LaTeX{}, with reference links displayed in their natural language forms.
A \textit{theorem} is associated with one or more proofs when available.
A \textit{proof} contains a title, contents, and references in the same format as a statement.
Finally, we collect \textit{other} pages (e.g. axioms, corollaries).
A \textit{reference} is a theorem, definition, or other page that is linked to within the contents of a statement or proof.
\autoref{fig:schema} shows the data format for theorems, definitions, and proofs in $\textsc{NaturalProofs}$.
All statements and the reference links connecting them form a \textit{reference graph}, shown in \autoref{fig:ref-graph}.
The reference graph can contain cycles, e.g. \texttt{Pythagoras's Theorem} and \texttt{Sum of Squares of Sine and Cosine} refer to each other in their proofs.

\textbf{Data sources and preprocessing.}
We describe how we retrieve data from each source and give an overview of preprocessing; for full details see Appendix \ref{apx:ssec:preprocess} and the Jupyter notebooks we release.

\begin{itemize}[itemsep=0cm,leftmargin=0.3cm,topsep=0cm]
\item \textbf{ProofWiki.}
We download the public ProofWiki XML dump,\footnote{\url{https://proofwiki.org/xmldump/latest.xml}. We use the November 12, 2020 version. ProofWiki is licensed under CC BY-SA 3.0.} which contains a snapshot of all pages on ProofWiki.
We filter pages according to manually designed rules (e.g. redirects, files, categories), and determine page type, title, contents, and references using each page's WikiMedia data structure.
\item \textbf{Stacks.}
We pull the Stacks GitHub repo,\footnote{\url{https://github.com/stacks/stacks-project}. We use the April 15, 2021 version (commit 4df67b8). Stacks is licensed under GNU Free Documentation License.} which contains multiple \LaTeX{} files for various sub-topics in algebraic geometry.
We extract statements and proofs by \LaTeX{} environment names.
For example, the content enclosed by \texttt{\textbackslash{}begin\{theorem\}} and \texttt{\textbackslash{}end\{theorem\}} would be considered a theorem.
\item \textbf{Textbooks.}
We searched for open-source math textbooks with rich theorem-proof structures and reference links.
Of those, we picked \textit{Introduction to Real Analysis}\footnote{\url{https://digitalcommons.trinity.edu/mono/7/}. Retrieved on April 15, 2021. We did not use the supplementary materials. This textbook is licensed under CC BY-NC-SA 3.0.} (\textbf{RA} in short) by William F. Trench and \textit{Elementary Number Theory: Primes, Congruences, and Secrets}\footnote{\url{https://github.com/williamstein/ent}. Retrieved on April 15, 2021. We provide a script to download and format the publicly available latex source.} (\textbf{NT} in short) by William Stein.
We downloaded the \LaTeX{} source of each textbook, and similarly extracted statements and proofs by environment names.
In both textbooks, every statement is either a theorem or a definition -- there are no statements that fall under "others".
\end{itemize}

\section{\textsc{NaturalProofs} Reference Retrieval and Generation Tasks}
\label{sec:task}

\begin{table}[t]
\footnotesize
\centering
\begin{tabular}{rr|rrr|rr}
\toprule
& \textbf{Split} & \textbf{P+S} & \textbf{ProofWiki} & \textbf{Stacks} & \textbf{RA} & \textbf{NT} \\
\toprule
\textbf{Examples} $|\mathcal{E}|$ & \textbf{total} & \textbf{25,271} & \textbf{14,698} & \textbf{10,573} & \textbf{167} & \textbf{40} \\
                                  & train & 21,446 & 12,424 & 9,022 & -- & -- \\
                                  & valid & 1,914 & 1,139 & 775 & -- & -- \\
                                  & test  & 1,911 & 1,135 & 776 & 167 & 40 \\
                                  \midrule
\textbf{Refs} $|\mathcal{R}|$     & train & 42,056 & 28,473 & 13,583 & -- & -- \\
                                  & valid & 45,805 & 30,671 & 15,134 & -- & -- \\
                                  & test  & 45,805 & 30,671 & 15,134 & 384 & 105 \\
                                  \midrule
\textbf{Refs/Ex} $|\yb|$          & train & 5.9 & 7.5 & 3.6 & -- & -- \\
                                  & valid & 5.6 & 7.5 & 2.9 & -- & -- \\
                                  & test  & 5.6 & 7.4 & 2.9 & 2.2 & 1.5 \\
\bottomrule
\addlinespace[0.4em]
\end{tabular}
\caption{
    \textsc{NaturalProofs} retrieval dataset statistics.
    \textbf{P+S} refers to the combined dataset from the ProofWiki and Stacks sources.
    \textbf{RA} (Real Analysis) and \textbf{NT} (Number Theory) are data from mathematical textbook sources that we use for zero-shot evaluation.
}
\label{tbl:dataset-retrieval-stats}
\end{table}

\textsc{NaturalProofs} opens many possible machine learning tasks that involve natural mathematical language.
We consider \textbf{mathematical reference retrieval}: given a theorem $\xb$, retrieve the set of references $\yb$ that occur in its proof.
An example is shown in \autoref{tbl:dataset-example}, where the task is to retrieve the underlined references given the title and contents of the theorem \texttt{Category of Monoids is Category}.
As a proof is ultimately written as an ordered collection of statements with references often occurring more than once, we also consider \textbf{mathematical reference generation}: generate the \textit{sequence} of references that occur in a given theorem's proof. 
These tasks represent a crucial aspect of theorem proving, in which a mathematician determines the key results that appear in a proof.

\myparagraph{Reference retrieval and generation}
Each theorem $\xb$ has a proof containing a sequence of references $\yb=(\rb_1,\ldots,\rb_{|\yb|})$, where each reference $\rb_m\in \mathcal{R}$ is either a theorem, definition, or other statement (see \S\ref{sec:dataset}).
We consider two tasks: \textit{retrieval} and \textit{generation}.

In the \textit{retrieval} task, given an input theorem $\xb$, a model assigns a score to each reference in $\mathcal{R}$, inducing a ranked list $\hat{\rb}^{(1)},\ldots,\hat{\rb}^{(|\mathcal{R}|)}$.
These ranked references are evaluated against the ground-truth reference set using standard retrieval metrics such as mean average precision ($\textsc{mAP}$), recall (\textsc{Rec}@$k$), and full recovery ($\textsc{Full}@k$), which checks whether all references in the proof are in the top-$k$ predicted rankings. 
This reflects the goal of fully proving a theorem using a fixed number of results.

In the \textit{generation} task, a model produces a variable-length sequence of references $(\hat{\rb}_1,\ldots,\hat{\rb}_{|\hat{\yb}|})$ given an input $\xb$, with the goal of exactly matching the ground-truth reference sequence $(\rb_1,\ldots,\rb_{|\yb|})$.
Unlike retrieval, generation requires the model to correctly predict the total number of references, the number of occurrences of each unique reference, and their orders in the proof.

\myparagraph{Input-output examples}
Using \textsc{NaturalProofs}, we derive examples of the  form $(\xb,\yb)$, where $\xb=(x_1,\ldots,x_T)$ is a theorem, and $\yb=(\rb_1,\ldots,\rb_{|\yb|})$ is the sequence of references that occur in the proof of $\xb$.
For retrieval, we transform each sequence into a set $\yb=\{\rb_1,\ldots,\rb_{|\yb|}\}$.
The set of all references, $\mathcal{R}$, consists of theorems, definitions, and other statements (see \S\ref{sec:dataset}).
We use theorems with at least one proof that has at least one reference, 
resulting in a dataset with roughly 25k examples and a reference set $\mathcal{R}$ with 46k unique references.
We partition the dataset into ProofWiki-only, Stacks-only, and textbook-only datasets.
\autoref{tbl:dataset-retrieval-stats} summarizes the size, total references, and average references per example in each dataset.

\myparagraph{Training and evaluation splits}
We design training and evaluation splits that reflect the real-world scenario of proving \textit{newly seen} theorems at evaluation time.
This requires careful attention, since 
naively sampling evaluation examples would yield evaluation theorems that appear as references in the training set.
To ensure that the theorems in the evaluation set have no overlap with the references in the training set,
we form an evaluation set using a randomly sampled subset of \textit{reference graph leaf nodes}, and use the remaining nodes as the training set (\autoref{fig:ref-graph}).
We use roughly half of the evaluation set for validation and the other half for testing.
Since evaluation theorems are not referred to in training examples, the reference set for training is smaller than that for evaluation (\autoref{tbl:dataset-retrieval-stats}).

\section{Methods}
\label{sec:methods}
As benchmark methods for our tasks, we introduce two \textit{parallel retrieval} methods, and a \textit{sequential retrieval} method trained for sequence generation. 
See Appendix~\ref{apx:experiments} for further implementation details.

\myparagraph{Parallel retrieval}
Given a theorem $\xb$, a retrieval model should assign high scores to references in the proof of $\xb$ and low scores to all other references, which corresponds to minimizing,
\begin{align}
\label{eqn:loss-exact}
    \mathcal{L}(\xb,\yb) &= \mathrm{KL}\left(p_*(\mathcal{R}|\xb)\|p_{\theta}(\mathcal{R}|\xb)\right)\\
    &\propto -\sum_{\rb\in \yb}\log \frac{\exp\left(s_\theta(\xb,\rb)\right)}{\sum_{\rb'\in\mathcal{R}}\exp\left(s_{\theta}(\xb,\rb')\right)} + \text{const},
\end{align}
where each distribution is over reference indices (i.e. in $\Delta^{(|\mathcal{R}|)}$),  and $p_*(\rb|\xb)\propto \mathbb{I}[\rb\in \yb]$. 
The denominator requires scores $s_{\theta}(\xb,\rb)$ for all $|\mathcal{R}|$ references, making backpropagation too expensive when a large-scale neural model is used to compute reference representations.
As a result we consider two variants: a \textit{pairwise} model that approximates \autoref{eqn:loss-exact}, and a \textit{joint} model that computes \autoref{eqn:loss-exact} but with implicit vector representations of each reference.

\myparagraph{Pairwise parameterization} 
This model contrasts each positive reference with a set of negatives,
\begin{align}
    \mathcal{L}(\xb, \rb, \yb_-) &= -\log \frac{\exp(s_\theta(\xb, \rb))}{\exp(s_\theta(\xb, \rb))+\sum_{\rb_-\in \yb_-}\exp(s_\theta(\xb, \rb_-))},
\end{align}
where $\rb$ is a reference that occurs in the proof of $\xb$, and $\yb_-$ is a (small) set of negative references.

We call this a pairwise parameterization since the score of each reference against the theorem x is computed independently of the other references,
$s_{\theta}(\xb,\rb)=f_{\theta_1}^{\text{thm}}(\xb)^\top g_{\theta_2}^{\text{ref}}(\rb)$.
This model represents retrieval methods such as the dense passage retriever \citep{karpukhin2020dense} and similar methods \citep{nogueira2020passage}, and allows for evaluating large-scale sequence models, in our case BERT~\citep{devlin2019bert}, on mathematical reference retrieval.

\myparagraph{Joint parameterization}
The second model scores all references in a single pass,
\begin{align}
    p_{\theta}(\mathcal{R}\mid \xb) &= \text{softmax}\left(\mathbf{R}f_{\theta}(\mathbf{x})\right),
\end{align}
where $\mathbf{R}\in \mathbb{R}^{|\mathcal{R}|\times d}$ is a reference embedding matrix and $f_{\theta}(\mathbf{x})\in \mathbb{R}^d$ is a neural theorem encoder.
This model allows for computing \autoref{eqn:loss-exact} exactly in our setting, 
but it must learn implicit representations of each reference, i.e. without observing reference contents.
To give the model access to representations that were learned using reference contents, we populate its embedding matrix as,
\begin{align}
\label{eqn:rmatrix}
    \mathbf{R}=\begin{bmatrix}
         \horzbar & g^{\text{ref}}(\rb_1) & \horzbar\\
         &\ldots &\\
         \horzbar & g^{\text{ref}}(\rb_{|\mathcal{R}|}) & \horzbar\\
\end{bmatrix},
\end{align} 
where $g^{\text{ref}}(\xb)$ is obtained by pretraining an independent model. 

\myparagraph{Sequential generation and retrieval}
Finally, we consider an autoregressive model,
\begin{align}
\label{eqn:autoreg}
    p_{\theta}(\rb_1,\ldots,\rb_{|\yb|}\mid\xb) &= \prod_{t=1}^{|\yb|+1}p_{\theta}(\rb_t|\rb_{<t}, \xb),
\end{align}
where $\rb_{|\yb|+1}$ is a special $\eos$ token denoting the end of the reference sequence.
The autoregressive model is trained to maximize the log-likelihood of ground-truth reference sequences.
Unlike the parallel retrieval models, this model predicts the order and total number of references and can predict multiple occurrences of each reference. It also adjusts its predictions based on preceding predictions.

For generation, a standard decoding algorithm (e.g. beam search) is used to generate a reference sequence $\hat{\yb}=(\hat{\rb}_1,\ldots,\hat{\rb}_{|\hat{\yb}|}\eos$).
For retrieval, we populate a ranked list using generations $\{\hat{\rb}_1,\ldots,\hat{\rb}_{|\hat{\yb}|}\}$ followed by references ordered according to the first step's probabilities, $p_{\theta}(\rb_{1}|\xb)$.

\section{Experiments}
\label{sec:experiments}
First, we benchmark the neural retrieval methods (\S\ref{sec:methods}) on mathematical reference retrieval in terms of their \textit{in-domain} performance (\autoref{tbl:retrieval-main}) and their \textit{out-of-domain} performance on an evaluation set formed from the textbooks in $\dsname$ (\autoref{tbl:retrieval-ood}).
We perform several analyses to better understand each method’s strengths, weaknesses, and the factors that contribute to their performance.

\myparagraph{In-domain performance} The BERT-based retrieval models show strong in-domain performance compared to the classical TF-IDF and naive baselines in terms of average precision, recall, and the ability to fully recover all true references within the top-$k$ results, as seen in \autoref{tbl:retrieval-main}.
On both ProofWiki and Stacks, the pairwise models outperform TF-IDF, with improvements that are consistent across reference types (Appendix~\autoref{tbl:retrieval-type}).

Joint parameterization substantially improves over the pairwise models that are the starting point of joint training.
On ProofWiki, the joint model ranks roughly 4 out of every 10 true references within its top 10 rankings (R@10 42.45) compared to 1 out of 10 for TF-IDF, and an impressive 75\% within its top 100.
For roughly half of the theorems, the joint model's top 100 references contain \textit{all} of the references 
needed to prove the theorem (Full@100 50.22).
On Stacks the recall@10 is similar at roughly 40\%, with a higher full recovery rate of 66\% for the top 100 results.

The gains from the joint parameterization are most prominent on ProofWiki, e.g. increasing mAP from 16.82 to 36.75.
Joint parameterization particularly excels at refining the top of the ranked list compared to pairwise parameterization; the percentage improvement in the @10 metrics are larger than those for @100 metrics.
On Stacks, the improvements are more modest: though mAP improves by 40\%, the other metrics are relatively close, suggesting that advances beyond the joint model are needed.
This demonstrates the importance of evaluating on multiple domains: each domain presents novel challenges for driving advances in modeling.
Finally, the BERT models trained on both ProofWiki and Stacks (\textbf{BERT (P+S)}) show the possibility of training a single multi-domain model, albeit with lower per-domain performance than the models trained individually on each domain.

\begin{table*}[t]
\setlength{\tabcolsep}{4pt}
\begin{center}
\resizebox{\linewidth}{!}{
\begin{tabular}{rr | rrrrr | rrrrr}
\toprule
& & \multicolumn{5}{c}{\textbf{ProofWiki}} & \multicolumn{5}{c}{\textbf{Stacks}} \\
\toprule
& & \textbf{mAP} & \textbf{R@10} & \textbf{R@100} & \textbf{Full@10} & \textbf{Full@100} & \textbf{mAP} &  \textbf{R@10} & \textbf{R@100} & \textbf{Full@10} & \textbf{Full@100} \\
\toprule
\multicolumn{2}{r|}{\textbf{Random}}     & 0.04 & 0.00 & 0.19 & 0.00 & 0.00 & 0.07 & 0.05 & 0.60 & 0.00 & 0.13 \\
\multicolumn{2}{r|}{\textbf{Frequency}}  & 3.38 & 5.90 & 24.30 & 0.44 & 2.29 & 0.91 & 1.76 & 11.27 & 0.13 & 2.45 \\
\multicolumn{2}{r|}{\textbf{TF-IDF}}     & 6.19 & 10.27 & 23.09 & 4.14 & 9.43 & 13.64 & 25.46 & 47.36 & 18.94 & 37.76 \\
\midrule
\multirow{2}{*}{\textbf{BERT (P+S)}} & +\textbf{pair} & 13.54 & 20.10 & 58.75 & 6.17 & 31.28 & 18.58 & 34.42 & 71.80 & 28.48 & 65.21 \\
& +\textbf{joint} & 32.71 & 37.59 & 73.72 & 17.71 & 48.90 & 26.88 & 35.71 & 72.68 & 28.99 & 66.11 \\
\multirow{2}{*}{\textbf{BERT (P/S)}} & +\textbf{pair} & 16.82 & 23.73 & 63.75 & 7.31 & 38.50 & 20.93 & 37.43 & \textbf{74.21} & 30.03 & \textbf{66.37} \\
& +\textbf{joint} & \textbf{36.75} & \textbf{42.45} & \textbf{75.90} & \textbf{20.35} & \textbf{50.22} & \textbf{28.32} & \textbf{39.10} & 73.61 & \textbf{31.96} & 65.59 \\
\bottomrule
\end{tabular}}
\caption{
    \textit{In-domain} performance on the mathematical reference retrieval task (test set).
    \textbf{BERT (P/S)} is finetuned on the part of dataset with the same source as the evaluation set, whereas \textbf{BERT (P+S)} is finetuned on the combined dataset from ProofWiki and Stacks sources.
    Recall is micro-averaged.
}
\label{tbl:retrieval-main}
\end{center}
\end{table*}

\begin{table*}[t]
\setlength{\tabcolsep}{6pt}
\begin{center}
\resizebox{\linewidth}{!}{
\begin{tabular}{llllll}
\toprule
\textbf{Source}&\multicolumn{2}{l}{\textbf{ProofWiki}} \\
\hline
\textbf{Theorem}&\multicolumn{2}{l}{\textbf{Category of Monoids is Category}} \\
&\multicolumn{2}{l}{Let $\mathrm{Mon}$ be the category of monoids.} \\
&\multicolumn{2}{l}{Then $\mathrm{Mon}$ is a metacategory.} \\
\hline
&\uline{\textbf{Ground-Truth Reference}}  & \uline{\textbf{Rank (Pairwise)}} & \uline{\textbf{Rank (Joint)}}  \\
&Metacategory  & 1 &1\\
&Identity Mapping is Left Identity & 4 & 5\\
&Identity Mapping is Right Identity  & 5 & 4\\
&Monoid  & 11 & 2 \\
&Composition of Mappings is Associative  & 21& 8 \\
&Identity Mapping is Automorphism & 117 & 64 \\
&Composite of Homomorphisms is Homomorphism& 261 & 54 \\
\hline
\textbf{\uline{Rank}}&\textbf{\uline{Reference (Pairwise)}}  &\multicolumn{2}{l}{\textbf{\uline{Reference (Joint)}}} \\ 
1&\textit{Metacategory }                            & \multicolumn{2}{l}{\textit{Metacategory}} \\
2&Monoid Category is Category                       & \multicolumn{2}{l}{\textit{Monoid}}  \\
3&Monoid Category                                   & \multicolumn{2}{l}{Identity Morphism} \\
4&\textit{Identity Mapping is Left Identity }       & \multicolumn{2}{l}{\textit{Identity Mapping is Right Identity}}   \\
5&\textit{Identity Mapping is Right Identity}       & \multicolumn{2}{l}{\textit{Identity Mapping is Left Identity}}  \\
6&Category                                          & \multicolumn{2}{l}{Associative}   \\
7&Composition of Morphisms                          & \multicolumn{2}{l}{Identity (Abstract Algebra)/Two-Sided Identity}   \\
8&Dual Category is Category                         & \multicolumn{2}{l}{\textit{Composition of Mappings is Associative}}   \\
9&Identity Morphism                                 & \multicolumn{2}{l}{Composition of Morphisms}   \\
10&Morphism Category                                & \multicolumn{2}{l}{Semigroup} \\
\bottomrule
\end{tabular}}
\caption{
    Retrieval for a representative theorem. Top: predicted ranks for ground-truth references using the pairwise (left) and its joint (right) BERT models. Bottom: top 10 retrievals from the pairwise (left) and joint (right) models.
    A retrieved reference is italicized when it is a ground-truth reference.
}
\label{tbl:prediction-example-representative}
\end{center}
\end{table*}
\myparagraph{Qualitative evaluation}
\autoref{tbl:prediction-example-representative} shows model predictions for a representative theorem, \texttt{Category of Monoids is Category}.
The pairwise model retrieves three out of seven true references within its top 50 results, while the joint model retrieves five out of seven.
The top 10 results for both models are comprised of references that are related to category theory, which is the subject of the theorem.
This illustrates the model's ability to retrieve \textit{relevant} references, while highlighting its inability to always perform the fine-grained distinction between a relevant reference and one that occurs in the ground-truth proof(s).
Arguably, such a system is still useful for providing hints to a user, so long as the user is confident that all of the true references are in a reasonably small set of results.

\begin{table}[t!]
\setlength{\tabcolsep}{4pt}
\begin{center}
\resizebox{0.65\linewidth}{!}{
\begin{tabular}{r | rrr | rrr}
\toprule
 & \multicolumn{3}{c}{\textbf{Real Analysis}} & \multicolumn{3}{c}{\textbf{Number Theory}} \\
\toprule
& \textbf{mAP} & \textbf{R@10} & \textbf{Full@10} & \textbf{mAP} & \textbf{R@10} & \textbf{Full@10}  \\
\midrule
\textbf{TF-IDF}     &  \textbf{15.79} & \textbf{34.65} & \textbf{27.54} & \textbf{16.42} & 39.62 & 30.00\\
\textbf{BERT-pair (P)}  &  13.24 & 24.01 & 19.16 & 15.12 & \textbf{41.51} & \textbf{35.00}\\
\textbf{+joint} & 11.24 & 20.97 & 16.77 &15.85 & 41.51 & 35.00\\
\textbf{BERT-pair (S)}&11.56 & 21.28 & 14.97 & 12.58 & 26.42 & 20.00 \\
\textbf{+joint} & 7.04 & 11.55& 9.58& 14.88 & 26.42 & 20.00\\

\bottomrule
\end{tabular}}
\end{center}
\caption{\textit{Zero-shot} retrieval performance on out-of-domain textbooks.
}
\label{tbl:retrieval-ood}
\end{table}

\myparagraph{Out-of-domain performance}
While strong in-domain performance drives applications in scenarios where training data is available, an ambitious goal is building a system with mathematical retrieval skills that automatically generalize to new resources.
To evaluate the retrieval methods in this zero-shot, out-of-domain setting, we use each textbook from $\dsname$ as an evaluation set.
This tests situations where the same theorem is expressed using different language (e.g. \autoref{tbl:dataset-example-same}), generalization across data formats, and whether retrieval ability from in-domain training transfers.

\autoref{tbl:retrieval-ood} shows the results. 
The pairwise BERT model trained on ProofWiki underperforms TF-IDF on the Real Analysis textbook, and has comparable performance on the Number Theory textbook.
Joint training did not improve out of domain performance, despite its favorable in-domain impact.
Training BERT on ProofWiki outperforms training on Stacks, showing that the training domain impacts out-of-domain generalization. 
ProofWiki's broad coverage of mathematics may help the model generalize better than the deep, single-topic coverage in Stacks.

The BERT models show some evidence of generalizing to out-of-domain mathematical sources, yet they do not show an advantage over traditional retrieval methods despite strong in-domain performance.
This aligns with recent findings about neural retrieval models in various zero-shot settings \citep{thakur2021beir}.
An exciting research direction is using $\dsname$ to develop and evaluate methods which improve not only in-domain performance, but out-of-domain generalization.

\begin{table}[t]
\setlength{\tabcolsep}{4pt}
\begin{center}
\small
\begin{tabular}{rrrrrrrrrrrr}
\toprule
& & \multicolumn{5}{c}{\textbf{Sequence}}& \multicolumn{2}{c}{\textbf{Multiset}} & \multicolumn{3}{c}{\textbf{Set}}\\
\cmidrule(lr){3-7}\cmidrule(lr){8-9}\cmidrule(lr){10-12}
& \textbf{Model} & \textbf{EM} & \textbf{Edit}($\downarrow$) & $\textbf{BLEU}_4$ & $\textbf{BLEU}_2$ & \textbf{Len} & \textbf{EM} & \textbf{F1} & \textbf{EM} & \textbf{F1} & $\textbf{BLEU}_1$  \\
\toprule
\multirow{5}{*}{\rotatebox[origin=c]{90}{\textbf{Stacks}}} & \textit{*-set} & 51.74 & 35.70 & 9.75 & 47.73 & 0.97 & 89.03 & 97.04 & 100.0 & 100.0 & 94.09\\
&\textit{*-multiset} & 49.42 & 38.13 & 9.71 & 47.71 & 1.00 & 100.0 & 100.0 & 100.0 & 100.0 & 100.0\\
&\textit{*-halfseq} & 0.00 & 70.49 & 6.13 & 12.08 & 0.30 & 0.00 & 56.86 & 0.65 & 58.01 & 16.87\\
\cmidrule(lr){2-12}
&Joint& 0.00 & 98.81 & 0.00 & \textbf{3.42} & 2.82 & 0.00 & \textbf{19.24} & 0.00 & \textbf{19.65} & \textbf{15.15}\\
& Autoregressive & \textbf{3.87} & \textbf{90.65} & 0.00 & 2.59 & \textbf{0.97} & \textbf{4.00} & 13.14 & \textbf{4.90} & 15.04 & 10.06\\
\hline
\multirow{5}{*}{\rotatebox[origin=c]{90}{\textbf{ProofWiki}}} &\textit{*-set} & 18.09 & 58.51 & 7.18 & 29.50 & 0.83 & 49.96 & 82.57 & 100.0 & 100.0 & 65.57\\
&\textit{*-multiset} & 19.23 & 58.09 & 16.68 & 52.89 & 1.00 & 100.0 & 100.0 & 100.0 & 100.0 & 100.0\\
&\textit{*-halfseq} & 0.00 & 58.84 & 25.88 & 29.17 & 0.41 & 0.00 & 63.33 & 4.21 & 70.26 & 30.55\\
\cmidrule(lr){2-12}
&Joint& 0.00 & 93.03 & 0.00 & 6.88 & 1.42 & 0.09 & 25.30 & 0.18 & \textbf{30.76} & 19.27\\
&Autoregressive& \textbf{3.69} & \textbf{84.30} & \textbf{5.48} & \textbf{11.90} & \textbf{1.18} & \textbf{3.78} & \textbf{25.61} & \textbf{4.65} & 28.97 & \textbf{20.81}\\
\bottomrule
\end{tabular}
\end{center}
\caption{
    In-domain \textit{generation} results.
    We show the autoregressive model, a retrieval-only baseline using the top-5 predictions from the joint retrieval model,
    and oracle benchmarks for correctly predicting the first half of the sequence (\textit{*-halfseq}), the full multiset with randomized order (\textit{*-multiset}), and the full set with randomized order (\textit{*-set)}.
    The best model-based method is  in bold.
}
\label{tbl:generation}
% \vspace{-1em}
\end{table}

\subsection{Reference Generation}
\label{ssec:refgen}
Next, we establish a benchmark for recovering the \textit{sequence} of references occurring in the proof of each theorem via the reference generation task (\S\ref{sec:task}).

\myparagraph{Metrics} We evaluate predicted reference sequences against ground-truth sequences using order-aware \textbf{sequence} metrics, as well as unordered \textbf{multiset} and \textbf{set}-based metrics. 
Sequence metrics include exact match (\textbf{EM}), edit-distance (\textbf{Edit}), standard $\textbf{BLEU}_4$ score which uniformly weights 1-4 gram precision, $\textbf{BLEU}_2$ with only 1-2 gram precision,
and average length ratio $\frac{\text{predicted}}{\text{true}}$ (\textbf{Len}).
Unordered metrics include exact match, \textbf{F1}-score (corpus level), and 1-gram precision $\textbf{BLEU}_1$.

\myparagraph{Methods} We use the autoregressive model to generate a reference sequence for each theorem using beam search.
As a retrieval-only baseline, we form a sequence using the joint retrieval model's top-5 predictions, ordered by retrieval score.
To judge performance and provide a benchmark for future work, we provide three oracle baselines: correctly predicting the first half of the sequence (\textit{*-halfseq}), the full multiset of references with random order (\textit{*-multiset}), and the set with random order (\textit{*-set}).

\myparagraph{Results} \autoref{tbl:generation} shows the in-domain generation results.
The task is challenging, with the autoregressive model exactly matching the ground-truth sequence roughly 3\% of the time.
The autoregressive model improves over the retrieval-only baseline on order-aware metrics, aside from $\textbf{BLEU}_2$ on Stacks.
It does length-prediction reasonably well, with length-ratios of 0.97 and 1.18, yet the multiset and set metrics indicate that the autoregressive model struggles to correctly predict the correct references, even after discarding order.
The oracle baselines indicate substantial room for future improvement-- for instance, predicting only half of each sequence correctly would move ProofWiki $\textbf{BLEU}_4$ from 5.48 to 25.88.
Developing models along the full spectrum from set-based retrieval, to reference generation, to full proof generation is an exciting use-case for $\dsname$.

\subsection{Ablation Studies}

\myparagraph{Initialization and autoregressive retrieval}
As shown in \autoref{tbl:autoregressive-ablation}, the autoregressive model trained for sequence generation substantially improves over the pairwise retrieval model, yet underperforms the joint model, which is trained specifically for retrieval.
Initializing the joint and autoregressive models using the pairwise model was necessary for achieving high performance; in particular, the reference information conveyed through the embedding matrix (\autoref{eqn:rmatrix}) was crucial.

\begin{table}[t]
\begin{minipage}[t]{.33\linewidth}
\setlength{\tabcolsep}{3pt}
\begin{center}
\footnotesize
\begin{tabular}[t]{ lcccc }
  \toprule
  \textbf{Init} & \textbf{Model} & \textbf{mAP} \\\toprule
  -- & Pairwise & 16.99\\
  \hline
  -- & Autoregressive & 17.77 \\ 
  $f^{\text{thm}}$ & Autoregressive & 25.07 \\ 
  $f^{\text{thm}},\mathbf{R}$ & Autoregressive & \textbf{35.37} \\
  \hline
  -- & Joint & 18.71 \\
  $f^{\text{thm}}$ & Joint & 28.95 \\
  $f^{\text{thm}},\mathbf{R}$ & Joint & \textbf{37.51} \\
  \bottomrule
\end{tabular}
\end{center}
\caption{Initializing with pairwise components,
and autoregressive retrieval (ProofWiki).
}
\label{tbl:autoregressive-ablation}
\end{minipage}
\hfill
\begin{minipage}[t]{.32\linewidth}
\setlength{\tabcolsep}{3pt}
\begin{center}
\footnotesize
\begin{tabular}[t]{cc|rr}
\toprule
\multicolumn{2}{c}{\textbf{Train}} & \multicolumn{2}{c}{\textbf{Eval}}\\
\textbf{Lang.} & \textbf{NatProof} & \textbf{PW} & \textbf{Stacks} \\
\toprule
\cmark & \xmark & 0.14 & 0.30 \\
\xmark & \cmark & 0.04 & 0.86 \\
\cmark & \cmark & \textbf{16.99} & \textbf{21.21} \\
\bottomrule
\end{tabular}
\end{center}
\caption{Language pretraining and $\dsname$ finetuning (pairwise retrieval, mAP).}
\label{tbl:retrieval-ablation-training}
\end{minipage}
\hfill
\begin{minipage}[t]{.33\linewidth}
\setlength{\tabcolsep}{3pt}
\begin{center}
\footnotesize
\begin{tabular}[t]{rccrr}
\toprule
& \textbf{Title} & \textbf{Content} & \textbf{PW} & \textbf{Stacks} \\
\toprule
\multirow{3}{*}{\rotatebox[origin=c]{90}{\textbf{TF-IDF}}} & \xmark & \cmark & 4.97 & 12.34 \\
                & \cmark & \xmark & \textbf{8.10} & 12.69 \\
                & \cmark & \cmark & 6.33 & \textbf{13.45} \\
\midrule
\multirow{3}{*}{\rotatebox[origin=c]{90}{\textbf{BERT}}}   & \xmark & \cmark & 16.19 & 19.12 \\
                & \cmark & \xmark & \textbf{24.48} & 19.15 \\
                & \cmark & \cmark & 16.99 & \textbf{21.21} \\
\bottomrule
\end{tabular}
\end{center}
\caption{
    Excluding (\xmark) the title or content of theorems and references (pairwise retrieval, mAP).
}
\label{tbl:retrieval-ablation-titlecontent}
\end{minipage}
\vspace{-2em}
\end{table}

\myparagraph{Language pretraining and \textsc{NaturalProofs} training}
The BERT model has two learning phases: pretraining on language data, and finetuning on \dsname.
As seen in \autoref{tbl:retrieval-ablation-training}, relying on language-pretraining alone without fine-tuning on $\dsname$ (top row) led to poor performance.
Conversely, training from scratch on $\dsname$ (middle row) was unsuccessful, suggesting that language pretraining served as an effective initialization for mathematical retrieval.

\myparagraph{Title and content ablation} 
Each theorem statement and reference consists of a title, as well as contents that is a mixture of symbolic mathematics and natural language. 
As seen in \autoref{tbl:retrieval-ablation-titlecontent}, ProofWiki's titles contain a large amount of useful information for retrieval-- TF-IDF and the pairwise BERT model performed better with only access to titles.
In principal, the title+content model could learn to ignore the contents if needed, so its lower performance shows a deficiency in the pairwise model.
On Stacks, the model performs best with both sources of information, though the degree of improvement suggests that leveraging the mathematical content remains as a fundamental challenge.

\section{Conclusion}
Building agents that understand and create mathematics using \textit{natural mathematical language} is a challenging research direction, providing a means for evaluating and developing machine learning methods capable of symbolic reasoning and natural language understanding.
As a step in this direction, we develop $\dsname$, a multi-domain dataset for studying mathematical reasoning in natural language.
$\dsname$ allows for evaluating \textit{in-domain} performance, and \textit{out-of-domain} generalization in broad and deep coverage mathematics, as well as real-world, low-resource settings.
We establish benchmarks for retrieval and generation tasks that represent key steps in real-world theorem proving,
and are tractable, yet challenging, for current large-scale neural sequence models.
$\dsname$ opens many promising avenues for future research.

\bibliographystyle{abbrvnat}
\bibliography{neurips_data_2021}

\begin{thebibliography}{47}
\providecommand{\natexlab}[1]{#1}
\providecommand{\url}[1]{\texttt{#1}}
\expandafter\ifx\csname urlstyle\endcsname\relax
  \providecommand{\doi}[1]{doi: #1}\else
  \providecommand{\doi}{doi: \begingroup \urlstyle{rm}\Url}\fi

\bibitem[Alemi et~al.(2016)Alemi, Chollet, Een, Irving, Szegedy, and
  Urban]{alemi2016deepmath}
A.~A. Alemi, F.~Chollet, N.~Een, G.~Irving, C.~Szegedy, and J.~Urban.
\newblock {DeepMath - Deep sequence models for premise selection}.
\newblock In \emph{Advances in Neural Information Processing Systems}, pages
  2243--2251, 2016.

\bibitem[Amini et~al.(2019)Amini, Gabriel, Lin, Koncel-Kedziorski, Choi, and
  Hajishirzi]{amini2019mathqa}
A.~Amini, S.~Gabriel, S.~Lin, R.~Koncel-Kedziorski, Y.~Choi, and H.~Hajishirzi.
\newblock {MathQA: Towards interpretable math word problem solving with
  operation-based formalisms}.
\newblock In \emph{NAACL HLT 2019 - 2019 Conference of the North American
  Chapter of the Association for Computational Linguistics: Human Language
  Technologies - Proceedings of the Conference}, volume~1, 2019.

\bibitem[Bansal et~al.(2019)Bansal, Loos, Rabe, Szegedy, and
  Wilcox]{bansal2019holist}
K.~Bansal, S.~Loos, M.~Rabe, C.~Szegedy, and S.~Wilcox.
\newblock {Holist: An environment for machine learning of higher-order theorem
  proving}.
\newblock In \emph{36th International Conference on Machine Learning, ICML
  2019}, volume 2019-June, 2019.

\bibitem[Brown et~al.(2020)Brown, Mann, Ryder, Subbiah, Kaplan, Dhariwal,
  Neelakantan, Shyam, Sastry, Askell, Agarwal, Herbert-Voss, Krueger, Henighan,
  Child, Ramesh, Ziegler, Wu, Winter, Hesse, Chen, Sigler, Litwin, Gray, Chess,
  Clark, Berner, McCandlish, Radford, Sutskever, and Amodei]{brown2020gpt3}
T.~Brown, B.~Mann, N.~Ryder, M.~Subbiah, J.~D. Kaplan, P.~Dhariwal,
  A.~Neelakantan, P.~Shyam, G.~Sastry, A.~Askell, S.~Agarwal, A.~Herbert-Voss,
  G.~Krueger, T.~Henighan, R.~Child, A.~Ramesh, D.~Ziegler, J.~Wu, C.~Winter,
  C.~Hesse, M.~Chen, E.~Sigler, M.~Litwin, S.~Gray, B.~Chess, J.~Clark,
  C.~Berner, S.~McCandlish, A.~Radford, I.~Sutskever, and D.~Amodei.
\newblock {Language Models are Few-Shot Learners}.
\newblock In H.~Larochelle, M.~Ranzato, R.~Hadsell, M.~F. Balcan, and H.~Lin,
  editors, \emph{Advances in Neural Information Processing Systems}, volume~33,
  pages 1877--1901. Curran Associates, Inc., 2020.
\newblock URL
  \url{https://proceedings.neurips.cc/paper/2020/file/1457c0d6bfcb4967418bfb8ac142f64a-Paper.pdf}.

\bibitem[Carter and Monks(2013)]{carter2013lurch}
N.~C. Carter and K.~G. Monks.
\newblock Lurch: a word processor that can grade students' proofs.
\newblock In C.~Lange, D.~Aspinall, J.~Carette, J.~H. Davenport, A.~Kohlhase,
  M.~Kohlhase, P.~Libbrecht, P.~Quaresma, F.~Rabe, P.~Sojka, I.~Whiteside, and
  W.~Windsteiger, editors, \emph{Joint Proceedings of the MathUI, OpenMath,
  {PLMMS} and ThEdu Workshops and Work in Progress at CICM, Bath, {UK}}, volume
  1010 of \emph{{CEUR} Workshop Proceedings}. CEUR-WS.org, 2013.
\newblock URL \url{http://ceur-ws.org/Vol-1010/paper-04.pdf}.

\bibitem[Clark et~al.(2020)Clark, Tafjord, and
  Richardson]{clark2020transformers}
P.~Clark, O.~Tafjord, and K.~Richardson.
\newblock {Transformers as Soft Reasoners over Language}.
\newblock In C.~Bessiere, editor, \emph{Proceedings of the Twenty-Ninth
  International Joint Conference on Artificial Intelligence, {IJCAI-20}}, pages
  3882--3890. International Joint Conferences on Artificial Intelligence
  Organization, 2020.

\bibitem[de~Moura et~al.(2015)de~Moura, Kong, Avigad, van Doorn, and von
  Raumer]{demoura2015lean}
L.~M. de~Moura, S.~Kong, J.~Avigad, F.~van Doorn, and J.~von Raumer.
\newblock The lean theorem prover (system description).
\newblock In A.~P. Felty and A.~Middeldorp, editors, \emph{CADE}, volume 9195
  of \emph{Lecture Notes in Computer Science}, pages 378--388. Springer, 2015.
\newblock ISBN 978-3-319-21400-9.
\newblock URL
  \url{http://dblp.uni-trier.de/db/conf/cade/cade2015.html#MouraKADR15}.

\bibitem[Devlin et~al.(2019)Devlin, Chang, Lee, and Toutanova]{devlin2019bert}
J.~Devlin, M.-W. Chang, K.~Lee, and K.~Toutanova.
\newblock {BERT}: Pre-training of deep bidirectional transformers for language
  understanding.
\newblock In \emph{Proceedings of the 2019 Conference of the North {A}merican
  Chapter of the Association for Computational Linguistics: Human Language
  Technologies, Volume 1 (Long and Short Papers)}, pages 4171--4186,
  Minneapolis, Minnesota, June 2019. Association for Computational Linguistics.
\newblock \doi{10.18653/v1/N19-1423}.
\newblock URL \url{https://www.aclweb.org/anthology/N19-1423}.

\bibitem[Ebrahimi et~al.(2021)Ebrahimi, Mager, Oncevay, Chaudhary, Chiruzzo,
  Fan, Ortega, Ramos, Rios, Vladimir, Giménez-Lugo, Mager, Neubig, Palmer,
  Solano, Vu, and Kann]{ebrahimi2021americasnli}
A.~Ebrahimi, M.~Mager, A.~Oncevay, V.~Chaudhary, L.~Chiruzzo, A.~Fan,
  J.~Ortega, R.~Ramos, A.~Rios, I.~Vladimir, G.~A. Giménez-Lugo, E.~Mager,
  G.~Neubig, A.~Palmer, R.~A.~C. Solano, N.~T. Vu, and K.~Kann.
\newblock Americasnli: Evaluating zero-shot natural language understanding of
  pretrained multilingual models in truly low-resource languages, 2021.

\bibitem[{European Organization For Nuclear Research} and
  {OpenAIRE}(2013)]{zenodo}
{European Organization For Nuclear Research} and {OpenAIRE}.
\newblock Zenodo, 2013.
\newblock URL \url{https://www.zenodo.org/}.

\bibitem[Ferreira and Freitas(2020{\natexlab{a}})]{ferreira2020natural}
D.~Ferreira and A.~Freitas.
\newblock Natural language premise selection: Finding supporting statements for
  mathematical text.
\newblock In \emph{Proceedings of the 12th Language Resources and Evaluation
  Conference}, pages 2175--2182, Marseille, France, May 2020{\natexlab{a}}.
  European Language Resources Association.
\newblock ISBN 979-10-95546-34-4.
\newblock URL \url{https://www.aclweb.org/anthology/2020.lrec-1.266}.

\bibitem[Ferreira and Freitas(2020{\natexlab{b}})]{ferreira2020premise}
D.~Ferreira and A.~Freitas.
\newblock Premise selection in natural language mathematical texts.
\newblock In \emph{Proceedings of the 58th Annual Meeting of the Association
  for Computational Linguistics}, pages 7365--7374, Online, July
  2020{\natexlab{b}}. Association for Computational Linguistics.
\newblock \doi{10.18653/v1/2020.acl-main.657}.
\newblock URL \url{https://www.aclweb.org/anthology/2020.acl-main.657}.

\bibitem[Gowers et~al.(2008)Gowers, Barrow-Green, and
  Leader]{gowers2008princetoncompanion}
T.~Gowers, J.~Barrow-Green, and I.~Leader.
\newblock \emph{The Princeton Companion to Mathematics}.
\newblock Princeton University Press, USA, illustrated edition edition, 2008.
\newblock ISBN 0691118809.

\bibitem[Hendrycks et~al.(2020)Hendrycks, Liu, Wallace, Dziedzic, Krishnan, and
  Song]{hendrycks2020pretrained}
D.~Hendrycks, X.~Liu, E.~Wallace, A.~Dziedzic, R.~Krishnan, and D.~Song.
\newblock Pretrained transformers improve out-of-distribution robustness, 2020.

\bibitem[Hendrycks et~al.(2021)Hendrycks, Burns, Kadavath, Arora, Basart, Tang,
  Song, and Steinhardt]{hendrycks2021measuring}
D.~Hendrycks, C.~Burns, S.~Kadavath, A.~Arora, S.~Basart, E.~Tang, D.~Song, and
  J.~Steinhardt.
\newblock Measuring mathematical problem solving with the math dataset, 2021.

\bibitem[Holland et~al.(2018)Holland, Hosny, Newman, Joseph, and
  Chmielinski]{holland2018dataset}
S.~Holland, A.~Hosny, S.~Newman, J.~Joseph, and K.~Chmielinski.
\newblock The dataset nutrition label: A framework to drive higher data quality
  standards.
\newblock \emph{arXiv preprint arXiv:1805.03677}, 2018.

\bibitem[Hu et~al.(2020)Hu, Ruder, Siddhant, Neubig, Firat, and
  Johnson]{hu2020xtreme}
J.~Hu, S.~Ruder, A.~Siddhant, G.~Neubig, O.~Firat, and M.~Johnson.
\newblock {XTREME}: A massively multilingual multi-task benchmark for
  evaluating cross-lingual generalisation.
\newblock In H.~D. III and A.~Singh, editors, \emph{Proceedings of the 37th
  International Conference on Machine Learning}, volume 119 of
  \emph{Proceedings of Machine Learning Research}, pages 4411--4421. PMLR,
  13--18 Jul 2020.
\newblock URL \url{http://proceedings.mlr.press/v119/hu20b.html}.

\bibitem[Huang et~al.(2019)Huang, Dhariwal, Song, and
  Sutskever]{huang2019gamepad}
D.~Huang, P.~Dhariwal, D.~Song, and I.~Sutskever.
\newblock Gamepad: A learning environment for theorem proving.
\newblock In \emph{International Conference on Learning Representations}, 2019.
\newblock URL \url{https://openreview.net/forum?id=r1xwKoR9Y7}.

\bibitem[Kang et~al.(2020)Kang, Head, Sidhu, Lo, Weld, and
  Hearst]{kang_document-level_2020}
D.~Kang, A.~Head, R.~Sidhu, K.~Lo, D.~Weld, and M.~A. Hearst.
\newblock Document-{Level} {Definition} {Detection} in {Scholarly} {Documents}:
  {Existing} {Models}, {Error} {Analyses}, and {Future} {Directions}.
\newblock In \emph{Proceedings of the {First} {Workshop} on {Scholarly}
  {Document} {Processing}}, pages 196--206, Online, Nov. 2020. Association for
  Computational Linguistics.
\newblock \doi{10.18653/v1/2020.sdp-1.22}.
\newblock URL \url{https://www.aclweb.org/anthology/2020.sdp-1.22}.

\bibitem[Karpukhin et~al.(2020)Karpukhin, O{\u{g}}uz, Min, Wu, Edunov, Chen,
  and Yih]{karpukhin2020dense}
V.~Karpukhin, B.~O{\u{g}}uz, S.~Min, L.~Wu, S.~Edunov, D.~Chen, and W.-t. Yih.
\newblock Dense passage retrieval for open-domain question answering.
\newblock \emph{arXiv preprint arXiv:2004.04906}, 2020.

\bibitem[Lample and Charton(2020)]{lample2020deep}
G.~Lample and F.~Charton.
\newblock Deep learning for symbolic mathematics.
\newblock In \emph{International Conference on Learning Representations}, 2020.
\newblock URL \url{https://openreview.net/forum?id=S1eZYeHFDS}.

\bibitem[{Le Bras} et~al.(2020){Le Bras}, Swayamdipta, Bhagavatula, Zellers,
  Peters, Sabharwal, and Choi]{lebras2020adversarial}
R.~{Le Bras}, S.~Swayamdipta, C.~Bhagavatula, R.~Zellers, M.~E. Peters,
  A.~Sabharwal, and Y.~Choi.
\newblock {Adversarial filters of dataset biases}, 2020.
\newblock ISSN 23318422.

\bibitem[Li et~al.(2021)Li, Yu, Wu, and Paulson]{li2021isarstep}
W.~Li, L.~Yu, Y.~Wu, and L.~C. Paulson.
\newblock Isarstep: a benchmark for high-level mathematical reasoning.
\newblock In \emph{International Conference on Learning Representations}, 2021.
\newblock URL \url{https://openreview.net/forum?id=Pzj6fzU6wkj}.

\bibitem[Ling et~al.(2017)Ling, Yogatama, Dyer, and Blunsom]{ling2017program}
W.~Ling, D.~Yogatama, C.~Dyer, and P.~Blunsom.
\newblock {Program induction by rationale generation: Learning to solve and
  explain algebraic word problems}.
\newblock In \emph{ACL 2017 - 55th Annual Meeting of the Association for
  Computational Linguistics, Proceedings of the Conference (Long Papers)},
  volume~1, 2017.
\newblock \doi{10.18653/v1/P17-1015}.

\bibitem[Megill and Wheeler(2019)]{megill2019metamath}
N.~D. Megill and D.~A. Wheeler.
\newblock \emph{Metamath: A Computer Language for Mathematical Proofs}.
\newblock Lulu Press, Morrisville, North Carolina, 2019.
\newblock {\tt http://us.metamath.org/downloads/metamath.pdf}.

\bibitem[Nogueira and Cho(2020)]{nogueira2020passage}
R.~Nogueira and K.~Cho.
\newblock Passage re-ranking with bert, 2020.

\bibitem[Nogueira~dos Santos et~al.(2020)Nogueira~dos Santos, Ma, Nallapati,
  Huang, and Xiang]{nogueira2020beyond}
C.~Nogueira~dos Santos, X.~Ma, R.~Nallapati, Z.~Huang, and B.~Xiang.
\newblock Beyond [{CLS}] through ranking by generation.
\newblock In \emph{Proceedings of the 2020 Conference on Empirical Methods in
  Natural Language Processing (EMNLP)}, pages 1722--1727, Online, Nov. 2020.
  Association for Computational Linguistics.
\newblock \doi{10.18653/v1/2020.emnlp-main.134}.
\newblock URL \url{https://www.aclweb.org/anthology/2020.emnlp-main.134}.

\bibitem[Petroni et~al.(2020)Petroni, Rockt{\"{a}}schel, Lewis, Bakhtin, Wu,
  Miller, and Riedel]{petroni2020language}
F.~Petroni, T.~Rockt{\"{a}}schel, P.~Lewis, A.~Bakhtin, Y.~Wu, A.~H. Miller,
  and S.~Riedel.
\newblock {Language models as knowledge bases?}
\newblock In \emph{EMNLP-IJCNLP 2019 - 2019 Conference on Empirical Methods in
  Natural Language Processing and 9th International Joint Conference on Natural
  Language Processing, Proceedings of the Conference}, 2020.
\newblock \doi{10.18653/v1/d19-1250}.

\bibitem[Polu and Sutskever(2020)]{polu2020generative}
S.~Polu and I.~Sutskever.
\newblock Generative language modeling for automated theorem proving, 2020.

\bibitem[Rabe et~al.(2021)Rabe, Lee, Bansal, and Szegedy]{rabe2021mathematical}
M.~N. Rabe, D.~Lee, K.~Bansal, and C.~Szegedy.
\newblock Mathematical reasoning via self-supervised skip-tree training.
\newblock In \emph{International Conference on Learning Representations}, 2021.
\newblock URL \url{https://openreview.net/forum?id=YmqAnY0CMEy}.

\bibitem[Radford et~al.(2019)Radford, Wu, Child, Luan, Amodei, and
  Sutskever]{radford2019language}
A.~Radford, J.~Wu, R.~Child, D.~Luan, D.~Amodei, and I.~Sutskever.
\newblock Language models are unsupervised multitask learners.
\newblock 2019.

\bibitem[Raffel et~al.(2020)Raffel, Shazeer, Roberts, Lee, Narang, Matena,
  Zhou, Li, and Liu]{raffel2020t5}
C.~Raffel, N.~Shazeer, A.~Roberts, K.~Lee, S.~Narang, M.~Matena, Y.~Zhou,
  W.~Li, and P.~J. Liu.
\newblock Exploring the limits of transfer learning with a unified text-to-text
  transformer.
\newblock \emph{Journal of Machine Learning Research}, 21\penalty0
  (140):\penalty0 1--67, 2020.
\newblock URL \url{http://jmlr.org/papers/v21/20-074.html}.

\bibitem[Rothe et~al.(2020)Rothe, Narayan, and Severyn]{rothe2020leveraging}
S.~Rothe, S.~Narayan, and A.~Severyn.
\newblock Leveraging pre-trained checkpoints for sequence generation tasks.
\newblock \emph{Transactions of the Association for Computational Linguistics},
  8:\penalty0 264--280, 2020.
\newblock \doi{10.1162/tacl_a_00313}.
\newblock URL \url{https://www.aclweb.org/anthology/2020.tacl-1.18}.

\bibitem[Roy and Roth(2015)]{roy2015solving}
S.~Roy and D.~Roth.
\newblock {Solving general arithmetic word problems}.
\newblock In \emph{Conference Proceedings - EMNLP 2015: Conference on Empirical
  Methods in Natural Language Processing}, 2015.
\newblock \doi{10.18653/v1/d15-1202}.

\bibitem[Saxton et~al.(2019)Saxton, Grefenstette, Hill, and
  Kohli]{saxton2018analysing}
D.~Saxton, E.~Grefenstette, F.~Hill, and P.~Kohli.
\newblock Analysing mathematical reasoning abilities of neural models.
\newblock In \emph{International Conference on Learning Representations}, 2019.
\newblock URL \url{https://openreview.net/forum?id=H1gR5iR5FX}.

\bibitem[Szegedy(2020)]{szegedy2020promising}
C.~Szegedy, editor.
\newblock \emph{A Promising Path Towards Autoformalization and General
  Artificial Intelligence}, 2020.

\bibitem[Tafjord et~al.(2020)Tafjord, Mishra, and
  Clark]{tafjord2020proofwriter}
O.~Tafjord, B.~D. Mishra, and P.~Clark.
\newblock Proofwriter: Generating implications, proofs, and abductive
  statements over natural language.
\newblock \emph{ArXiv}, abs/2012.13048, 2020.

\bibitem[Thakur et~al.(2021)Thakur, Reimers, Rücklé, Srivastava, and
  Gurevych]{thakur2021beir}
N.~Thakur, N.~Reimers, A.~Rücklé, A.~Srivastava, and I.~Gurevych.
\newblock Beir: A heterogenous benchmark for zero-shot evaluation of
  information retrieval models.
\newblock \emph{arXiv preprint arXiv:2104.08663}, 4 2021.
\newblock URL \url{https://arxiv.org/abs/2104.08663}.

\bibitem[Thurston(1994)]{thurston1994proof}
W.~P. Thurston.
\newblock On proof and progress in mathematics.
\newblock \emph{arXiv:math/9404236}, Mar. 1994.
\newblock URL \url{http://arxiv.org/abs/math/9404236}.
\newblock arXiv: math/9404236.

\bibitem[Urban(2006)]{urban2006mptp}
J.~Urban.
\newblock Mptp 0.2: Design, implementation, and initial experiments.
\newblock \emph{J. Autom. Reason.}, 37\penalty0 (1–2):\penalty0 21–43, Aug.
  2006.
\newblock ISSN 0168-7433.
\newblock \doi{10.1007/s10817-006-9032-3}.
\newblock URL \url{https://doi.org/10.1007/s10817-006-9032-3}.

\bibitem[Wang and Deng(2020)]{wang2020learning}
M.~Wang and J.~Deng.
\newblock {Learning to Prove Theorems by Learning to Generate Theorems}.
\newblock In H.~Larochelle, M.~Ranzato, R.~Hadsell, M.~F. Balcan, and H.~Lin,
  editors, \emph{Advances in Neural Information Processing Systems}, volume~33,
  pages 18146--18157. Curran Associates, Inc., 2020.
\newblock URL
  \url{https://proceedings.neurips.cc/paper/2020/file/d2a27e83d429f0dcae6b937cf440aeb1-Paper.pdf}.

\bibitem[Wang et~al.(2020)Wang, Brown, Kaliszyk, and
  Urban]{wang2020exploration}
Q.~Wang, C.~Brown, C.~Kaliszyk, and J.~Urban.
\newblock {Exploration of neural machine translation in autoformalization of
  mathematics in Mizar}.
\newblock In \emph{CPP 2020 - Proceedings of the 9th ACM SIGPLAN International
  Conference on Certified Programs and Proofs, co-located with POPL 2020},
  2020.
\newblock \doi{10.1145/3372885.3373827}.

\bibitem[Whalen(2016)]{whalen2016holophrasm}
D.~Whalen.
\newblock Holophrasm: a neural automated theorem prover for higher-order logic,
  2016.

\bibitem[Williams et~al.(2018)Williams, Nangia, and Bowman]{williams2018mnli}
A.~Williams, N.~Nangia, and S.~R. Bowman.
\newblock {A broad-coverage challenge corpus for sentence understanding through
  inference}.
\newblock In \emph{NAACL HLT 2018 - 2018 Conference of the North American
  Chapter of the Association for Computational Linguistics: Human Language
  Technologies - Proceedings of the Conference}, 2018.
\newblock ISBN 9781948087278.
\newblock \doi{10.18653/v1/n18-1101}.

\bibitem[Wolf et~al.(2020)Wolf, Debut, Sanh, Chaumond, Delangue, Moi, Cistac,
  Rault, Louf, Funtowicz, Davison, Shleifer, von Platen, Ma, Jernite, Plu, Xu,
  Scao, Gugger, Drame, Lhoest, and Rush]{wolf2020transformers}
T.~Wolf, L.~Debut, V.~Sanh, J.~Chaumond, C.~Delangue, A.~Moi, P.~Cistac,
  T.~Rault, R.~Louf, M.~Funtowicz, J.~Davison, S.~Shleifer, P.~von Platen,
  C.~Ma, Y.~Jernite, J.~Plu, C.~Xu, T.~L. Scao, S.~Gugger, M.~Drame, Q.~Lhoest,
  and A.~M. Rush.
\newblock Transformers: State-of-the-art natural language processing.
\newblock In \emph{Proceedings of the 2020 Conference on Empirical Methods in
  Natural Language Processing: System Demonstrations}, pages 38--45, Online,
  Oct. 2020. Association for Computational Linguistics.
\newblock URL \url{https://www.aclweb.org/anthology/2020.emnlp-demos.6}.

\bibitem[Wu et~al.(2021)Wu, Rabe, Li, Ba, Grosse, and Szegedy]{wu2021lime}
Y.~Wu, M.~Rabe, W.~Li, J.~Ba, R.~Grosse, and C.~Szegedy.
\newblock Lime: Learning inductive bias for primitives of mathematical
  reasoning, 2021.

\bibitem[Yang and Deng(2019)]{yang2019learning}
K.~Yang and J.~Deng.
\newblock {Learning to prove theorems via interacting with proof assistants}.
\newblock In \emph{36th International Conference on Machine Learning, ICML
  2019}, volume 2019-June, 2019.

\end{thebibliography}

\section*{Checklist}

\begin{enumerate}

\item For all authors...
\begin{enumerate}
  \item Do the main claims made in the abstract and introduction accurately reflect the paper's contributions and scope?
    \answerYes{}
  \item Did you describe the limitations of your work?
    \answerYes{We discussed limitations throughout our experimental analysis.}
  \item Did you discuss any potential negative societal impacts of your work?
    \answerNA{Our work pertains to use of natural language in mathematical theorem proving, and more generally reasoning in artificial intelligence. 
    Although a general reasoning agent may present negative societal impacts, we do not foresee any immediate negative societal impact from the domain, dataset, tasks, and study that we present here.
    Instead, we foresee positive societal impacts through education and scientific discovery from building systems that understand and create natural mathematical content.}
  \item Have you read the ethics review guidelines and ensured that your paper conforms to them?
    \answerYes{}
\end{enumerate}

\item If you are including theoretical results...
\begin{enumerate}
  \item Did you state the full set of assumptions of all theoretical results?
    \answerNA{We did not include theoretical results.}
	\item Did you include complete proofs of all theoretical results?
    \answerNA{}
\end{enumerate}

\item If you ran experiments (e.g. for benchmarks)...
\begin{enumerate}
  \item Did you include the code, data, and instructions needed to reproduce the main experimental results (either in the supplemental material or as a URL)?
    \answerYes{We released our code as a GitHub repo and our dataset on Zenodo.}
  \item Did you specify all the training details (e.g., data splits, hyperparameters, how they were chosen)?
    \answerYes{We specified data splits in \autoref{sec:task}, and hyperparameters in \autoref{apx:experiments}.}
	\item Did you report error bars (e.g., with respect to the random seed after running experiments multiple times)?
    \answerNo{We report results from a single run of each experiment due to computational constraints.}
	\item Did you include the total amount of compute and the type of resources used (e.g., type of GPUs, internal cluster, or cloud provider)?
    \answerYes{We specified the computing resources in \autoref{apx:experiments}.}
\end{enumerate}

\item If you are using existing assets (e.g., code, data, models) or curating/releasing new assets...
\begin{enumerate}
  \item If your work uses existing assets, did you cite the creators?
    \answerYes{In \autoref{sec:dataset}, we cited the authors of mathematical textbooks we used as data sources. ProofWiki and Stacks are collaboratively created on the web.}
  \item Did you mention the license of the assets?
    \answerYes{We noted the license of each data source in \autoref{sec:dataset}, and verified that all permit redistribution with modification for non-commercial purposes.}
  \item Did you include any new assets either in the supplemental material or as a URL?
    \answerYes{We released the \textsc{NaturalProofs} dataset on Zenodo, and provide additional resources in a public Github repository.}
  \item Did you discuss whether and how consent was obtained from people whose data you're using/curating?
    \answerNA{The licenses of the data indicate that our usage is permitted.}
  \item Did you discuss whether the data you are using/curating contains personally identifiable information or offensive content?
    \answerNA{The data we are using/curating contains no PII or offensive content.}
\end{enumerate}

\item If you used crowdsourcing or conducted research with human subjects...
\begin{enumerate}
  \item Did you include the full text of instructions given to participants and screenshots, if applicable?
    \answerNA{We did not use crowdsourcing or conduct research with human subjects.}
  \item Did you describe any potential participant risks, with links to Institutional Review Board (IRB) approvals, if applicable?
    \answerNA{}
  \item Did you include the estimated hourly wage paid to participants and the total amount spent on participant compensation?
    \answerNA{}
\end{enumerate}

\end{enumerate}

\clearpage
\appendix

\section*{Appendix}

\section{Dataset Details}

\autoref{tbl:dataset-example-more} shows example theorems and proofs from more data sources.
\autoref{tbl:dataset-example-same} shows an example of the same theorem extracted from different sources.
\autoref{tbl:dataset-stats-detail} gives more detailed statistics of the dataset.
\autoref{json-example} shows the JSON format of an example theorem, whereas \autoref{fig:schema} shows the data schema we use to standardize data collected from different sources.

\begin{table}[h]
\footnotesize
\setlength{\tabcolsep}{2pt}
\centering
\begin{tabular}{l p{12cm}}
\toprule
\textbf{Source} & \textbf{Stacks} \\
\hline
\textbf{Theorem} & \textbf{Lemma 9.7} \\
& Let $S$ be a scheme. Let $f : X \to S$ be locally of finite type with $X$ quasi-compact. Then $\text{size}(X) \leq \text{size}(S)$. \\
\hline
\textbf{Proof}
& We can find a finite affine open covering $X = \bigcup_{i = 1, \ldots n} U_i$ such that each $U_i$ maps into an affine open $S_i$ of $S$. Thus by \uline{Lemma 9.5} we reduce to the case where both $S$ and $X$ are affine. In this case by \uline{Lemma 9.4} we see that it suffices to show \\
& $|A[x_1, \ldots, x_n]| \leq \max\{\aleph_0, |A|\}.$ \\ 
& We omit the proof of this inequality. \\
\bottomrule
\addlinespace[0.2em]
\toprule
\textbf{Source} & \textbf{Textbook: Number Theory} \\
\hline
\textbf{Theorem} & \textbf{Proposition 2.1.13} \\
& If $\gcd(a,n)=1$, then the equation $ax \equiv b \pmod{n}$ has a solution, and that solution is unique modulo~$n$. \\
\hline
\textbf{Proof}
& Let~$R$ be a complete set of residues modulo~$n$, so there is a unique element of~$R$ that is congruent to~$b$ modulo~$n$. \\
& By \uline{Lemma 2.1.12}, $aR$ is also a complete set of residues modulo~$n$, so there is a unique element $ax \in aR$ that is congruent to~$b$ modulo~$n$, and we have $ax \equiv b \pmod{n}$. \\
\bottomrule
\addlinespace[0.2em]
\end{tabular}
\caption{
    Example theorems and their proofs from the Stacks and Number Theory textbook sources.
}
\label{tbl:dataset-example-more}
\end{table}

\begin{table}[h]
\footnotesize
\setlength{\tabcolsep}{2pt}
\centering
\begin{tabular}{l p{12cm}}
\toprule
\textbf{Source} & \textbf{ProofWiki} \\
\hline
\textbf{Theorem} & \textbf{Solution of Linear Congruence/Unique iff Coprime to Modulus} \\
& If $\gcd \{a, n\} = 1$, then $a x \equiv b \pmod n$ has a \uline{unique} solution. \\
\hline
\textbf{Proof}
& From \uline{Solution of Linear Congruence: Existence}: \\
& the problem of finding all integers satisfying the \uline{linear congruence} $a x \equiv b \pmod n$ \\
& is the same problem as: \\
& the problem of finding all the $x$ values in the \uline{linear Diophantine equation} $a x - n y = b$. \\
& Let: $\gcd \{a, n\} = 1$ \\
& Let $x = x_0, y = y_0$ be one solution to the \uline{linear Diophantine equation}: $a x - n y = b$ \\
& From \uline{Solution of Linear Diophantine Equation}, the general solution is: \\
& $\forall k \in \mathbb{Z}: x = x_0 + n k, y = y_0 + a k$ \\
& But: $\forall k \in \mathbb{Z}: x_0 + n k \equiv x_0 \pmod n$ \\
& Hence $x \equiv x_0 \pmod n$ is the only solution of $a x \equiv b \pmod n$. \\
\bottomrule
\toprule
\textbf{Source} & \textbf{Textbook: Number Theory} \\
\hline
\textbf{Theorem} & \textbf{Units} \\
& If $\gcd(a,n)=1$, then the equation $ax \equiv b \pmod{n}$ has a solution, and that solution is unique modulo~$n$. \\
\hline
\textbf{Proof}
& Let~$R$ be a complete set of residues modulo~$n$, so there is a unique element of~$R$ that is congruent to~$b$ modulo~$n$. \\
& By \uline{Lemma 2.1.12}, $aR$ is also a complete set of residues modulo~$n$, so there is a unique element $ax \in aR$ that is congruent to~$b$ modulo~$n$, and we have $ax \equiv b \pmod{n}$. \\
\bottomrule
\addlinespace[0.4em]
\end{tabular}
\caption{
    Example of the same theorem extracted from two different sources.
}
\label{tbl:dataset-example-same}
\end{table}

\begin{table*}[ht]
\setlength{\tabcolsep}{3pt}
\centering
\resizebox{\columnwidth}{!}{
\begin{tabular}{rr | l l@{\hspace{1pt}}l@{\hspace{1pt}}l | l l@{\hspace{1pt}}l@{\hspace{1pt}}l | l l@{\hspace{1pt}}l@{\hspace{1pt}}l | l l@{\hspace{1pt}}l@{\hspace{1pt}}l | l l@{\hspace{1pt}}l@{\hspace{1pt}}l}
\toprule
& \textbf{Source} & \multicolumn{3}{c}{\textbf{All}} & & \multicolumn{3}{c}{\textbf{ProofWiki}} & & \multicolumn{3}{c}{\textbf{Stacks}} & & \multicolumn{3}{c}{\textbf{Textbook: RA}} & & \multicolumn{3}{c}{\textbf{Textbook: NT}} \\
\toprule
\textbf{Type} & \textbf{Attr} & \textbf{mean} & \textbf{25\%} & \textbf{50\%} & \textbf{75\%} & \textbf{mean} & \textbf{25\%} & \textbf{50\%} & \textbf{75\%} & \textbf{mean} & \textbf{25\%} & \textbf{50\%} & \textbf{75\%} & \textbf{mean} & \textbf{25\%} & \textbf{50\%} & \textbf{75\%} & \textbf{mean} & \textbf{25\%} & \textbf{50\%} & \textbf{75\%} \\
\toprule
\multirow{5}{*}{\rotatebox[origin=c]{90}{\textbf{Theorem}}}
 & N       & \textit{32,579} & - & - & -    & \textit{19,734} & - & - & -      & \textit{12,479} & - & - & -   & \textit{298} & - & - & -    & \textit{68} & - & - & - \\
 & Chars   & 320.0 & 146 & 275 & 433        & 277.9 & 93 & 238 & 393           & 388.6 & 215 & 331 & 491       & 278.2 & 152 & 225 & 355     & 158.4 & 98 & 140 & 179 \\
 & Tokens  & 46.7 & 21 & 39 & 63            & 38.2 & 14 & 32 & 53              & 60.6 & 35 & 52 & 76           & 33.6 & 19 & 29 & 41         & 23.7 & 14 & 21 & 30 \\
 & Lines   & 5.9 & 2 & 4 & 8                & 3.6 & 1 & 3 & 5                  & 9.7 & 4 & 8 & 12              & 8.4 & 4 & 7 & 11            & 4.5 & 2 & 4 & 5 \\
 & Refs    & 1.8 & 0 & 0 & 3                & 2.8 & 0 & 3 & 4                  & 0.2 & 0 & 0 & 0               & 0.0 & 0 & 0 & 0             & 0.0 & 0 & 0 & 0 \\
 \midrule
\multirow{5}{*}{\rotatebox[origin=c]{90}{\textbf{Proof}}}
 & N       & \textit{32,012} & - & - & -    & \textit{19,234} & - & - & -      & \textit{12,479} & - & - & -   & \textit{235} & - & - & -    & \textit{64} & - & - & - \\
 & Chars   & 1,123.8 & 388 & 770 & 1,449    & 1,170.0 & 444 & 810 & 1,470      & 1,053.1 & 280 & 705 & 1,422   & 1231.0 & 442 & 876 & 1,634  & 655.7 & 327 & 551 & 732 \\
 & Tokens  & 181.5 & 57 & 121 & 236         & 199.3 & 68 & 134 & 254           & 155.5 & 36 & 101 & 211        & 128.9 & 50 & 92 & 165       & 97.2 & 47 & 87 & 115 \\
 & Lines   & 24.9 & 8 & 16 & 32             & 25.8 & 9 & 18 & 33               & 23.4 & 6 & 15 & 31            & 36.1 & 14 & 27 & 47         & 16.1 & 8 & 13 & 18 \\
 & Refs    & 5.6 & 2 & 3 & 7                & 7.4 & 2 & 5 & 9                  & 3.0 & 1 & 2 & 4               & 1.6 & 0 & 1 & 2             & 0.9 & 0 & 1 & 1 \\
 \midrule
\multirow{5}{*}{\rotatebox[origin=c]{90}{\textbf{Definition}}}
 & N       & \textit{14,230} & - & - & -    & \textit{12,420} & - & - & -      & \textit{1,687} & - & - & -    & \textit{86} & - & - & -     & \textit{37} & - & - & - \\
 & Chars   & 362.3 & 152 & 300 & 491        & 349.3 & 131 & 289 & 478          & 459.0 & 251 & 380 & 577       & 411.8 & 246 & 356 & 509     & 199.5 & 118 & 159 & 262 \\
 & Tokens  & 48.4 & 18 & 39 & 65            & 45.0 & 15 & 35 & 61              & 73.2 & 41 & 61 & 91           & 58.6 & 33 & 49 & 74         & 32.6 & 21 & 28 & 43 \\
 & Lines   & 5.0 & 1 & 4 & 6                & 4.2 & 1 & 3 & 6                  & 10.7 & 5 & 9 & 13             & 13.3 & 8 & 11 & 17          & 5.1 & 3 & 4 & 7 \\
 & Refs    & 2.9 & 0 & 2 & 4                & 3.3 & 1 & 3 & 5                  & 0.4 & 0 & 0 & 1               & 0.0 & 0 & 0 & 0             & 0.0 & 0 & 0 & 0 \\
 \midrule
\multirow{5}{*}{\rotatebox[origin=c]{90}{\textbf{Other}}}
 & N       & \textit{1,974} & - & - & -     & \textit{1,006} & - & - & -       & \textit{968} & - & - & - \\
 & Chars   & 1,399.8 & 712 & 1,109 & 1,680  & 1,836.5 & 1,018 & 1,431 & 2,131  & 945.9 & 480 & 802 & 1,198 \\
 & Tokens  & 212.1 & 101 & 158 & 250        & 286.1 & 145 & 206 & 337          & 135.2 & 70 & 113 & 168 \\
 & Lines   & 34.4 & 18 & 28 & 42            & 46.7 & 28 & 39 & 49              & 21.7 & 10 & 18 & 27 \\
 & Refs    & 5.7 & 1 & 3 & 7                & 9.2 & 4 & 7 & 11                 & 2.0 & 0 & 1 & 3 \\
\bottomrule
\end{tabular}
}
\caption{
    \textsc{NaturalProofs} dataset statistics (detailed).
}
\label{tbl:dataset-stats-detail}
\end{table*}
\begin{figure*}[ht]
\begin{minted}[
               linenos=false,
               xleftmargin=2pt,
               tabsize=2,
               fontsize=\scriptsize,
               ]{js}
{
  "id": 5480,
  "type": "theorem",
  "label": "Category of Monoids is Category",
  "categories": [ "Category of Monoids" ],
  "toplevel_categories": [ "Algebra", "Set Theory", "Abstract Algebra", "Category Theory" ],
  "recursive_categories": [
    "Category Theory",
    "Algebra",
    "Abstract Algebra",
    "Category of Monoids",
    "Set Theory",
    "Examples of Categories"
  ],
  "title": "Category of Monoids is Category",
  "contents": [
    "Let $\\mathbf{Mon}$ be the [[Definition:Category of Monoids|category of monoids]].",
    "Then $\\mathbf{Mon}$ is a [[Definition:Metacategory|metacategory]]."
  ],
  "refs": [
    "Definition:Category of Monoids",
    "Definition:Metacategory"
  ],
  "ref_ids": [ 22919, 21454 ],
  "proofs": [
    {
      "contents": [
        "Let us verify the axioms $(C1)$ up to $(C3)$ for a [[Definition:Metacategory|metacategory]].",
        "We have [[Composite of Homomorphisms on Algebraic Structure is Homomorphism]], verifying $(C1)$.",
        "We have [[Identity Mapping is Automorphism]] providing $\\operatorname{id}_S$ for every 
            [[Definition:Monoid|monoid]] $\\left({S, \\circ}\\right)$.",
        "Now, $(C2)$ follows from [[Identity Mapping is Left Identity]] and 
            [[Identity Mapping is Right Identity]].",
        "Finally, $(C3)$ follows from [[Composition of Mappings is Associative]].",
        "Hence $\\mathbf{Mon}$ is a [[Definition:Metacategory|metacategory]].",
        "{{qed}}",
        "[[Category:Category of Monoids]]",
        "sppgcr1pruam0jkf2euhyvt6y3jpnt0"
      ],
      "refs": [
        "Definition:Metacategory",
        "Composite of Homomorphisms is Homomorphism/Algebraic Structure",
        "Identity Mapping is Automorphism",
        "Definition:Monoid",
        "Identity Mapping is Left Identity",
        "Identity Mapping is Right Identity",
        "Composition of Mappings is Associative",
        "Definition:Metacategory"
      ],
      "ref_ids": [ 21454, 3852, 418, 19948, 217, 4387, 1494, 21454 ]
    }
  ]
}
\end{minted}
\caption{
    \textsc{NaturalProofs} JSON for the theorem and proof shown in \autoref{tbl:dataset-example}.
    Using the notation in \autoref{sec:task}, an $(\xb,\yb)$ example is formed where $\xb$ is the concatenation of \mintinline{js}{'title'} and \mintinline{js}{'contents'}, and $\yb$ is a set formed with \mintinline{js}{'ref_ids'} of one of the proofs.
} 
\label{json-example}
\end{figure*}
\begin{figure*}[ht]
\begin{minipage}[t]{0.48\textwidth}
\begin{minted}[
               linenos=false,
               xleftmargin=2pt,
               tabsize=2,
               fontsize=\scriptsize,
               ]{js}
Dataset: {
  'dataset': {
    'theorems': [Statement],
    'definitions': [Statement],
    'others': [Statement],
    'retrieval_examples': [int], // deprecated
  },
  'splits': {
    'train': {
      'ref_ids': [int],
      'examples': [(int, int)],
      // pairs of theorem id and index of proof
    },
    'valid': {
      'ref_ids': [int],
      'examples': [(int, int)],
    },
    'test': {
      'ref_ids': [int],
      'examples': [(int, int)],
    },
  },
}
\end{minted}
\end{minipage}
\begin{minipage}[t]{0.48\textwidth}
\begin{minted}[
               linenos=false,
               xleftmargin=2pt,
               tabsize=2,
               fontsize=\scriptsize,
               ]{js}
Statement: {
  'id': int,
  'type': string,
  'label': string,
  'categories': [string],
  'toplevel_categories': [string], // ProofWiki only
  'recursive_categories': [string], // ProofWiki only
  'title': string,
  'contents': [string],
  'refs': [string],
  'ref_ids': [int],
  'proofs': [Proof], // for theorems only
}
\end{minted}
\vspace{0.5em}
\begin{minted}[
               linenos=false,
               xleftmargin=2pt,
               tabsize=2,
               fontsize=\scriptsize,
               ]{js}
Proof: {
  'contents': [string],
  'refs': [string],
  'ref_ids': [int],
}
\end{minted}
\end{minipage}
\captionof{figure}{
    \textsc{NaturalProofs} dataset schema.
}
\label{fig:schema}
\end{figure*}

\subsection{Preprocessing Details}
\label{apx:ssec:preprocess}

\textbf{ProofWiki.}
The theorem, definition, and proof contents are contained in a WikiMedia section that is determined for each page type according to a hand-defined rule.
Since the roughly 1,000 other pages have varying page structures, we use their entire contents instead of a single section's contents.
In addition to well-formed axiom and corollary statements, the other pages include misformatted theorem or definition statements that occur as references elsewhere in the corpus.

\textbf{Stacks and textbooks.}
The raw data we obtain from Stacks and textbook sources are \LaTeX{} source code.
For each data source, we look up with a pre-defined list of environment names, and parse the contents enclosed in these environments into statements or proofs.
Each proof is associated with the environment that immediately precedes it.
As a result, each theorem has at most one proof.
\autoref{tbl:preprocess} lists the mapping from \LaTeX{} environment name to the data type in the \textsc{NaturalProofs} taxonomy.

A few misc notes:
\begin{itemize}
\item In Stacks, statements do not have titles, but each has a label with semantic meaning (e.g. \texttt{sets-lemma-bound-finite-type} for the example in \autoref{tbl:dataset-example-more}), so we use it as a pseudo-title.
\item In the Number Theory textbook, proofs are bounded by (\texttt{\textbackslash{}proof}, \texttt{\textbackslash{}bbox}) instead of (\texttt{\textbackslash{}begin\{proof\}}, \texttt{\textbackslash{}end\{proof\}}).
\end{itemize}

\begin{table}[h]
\begin{minipage}[t]{.32\linewidth}
\centering
\begin{tabular}{l|l}
\toprule
\textbf{Source} & Stacks \\
\toprule
\textbf{\LaTeX{} env} & \textbf{Type} \\
\midrule
theorem & theorem \\
lemma & theorem \\
proposition & theorem \\
definition & definition \\
remark & other \\
remarks & other \\
proof & proof \\
\bottomrule
\end{tabular}
\end{minipage}
\hfill
\begin{minipage}[t]{.32\linewidth}
\centering
\begin{tabular}{l|l}
\toprule
\textbf{Source} & Textbook: RA \\
\toprule
\textbf{\LaTeX{} env} & \textbf{Type} \\
\midrule
theorem & theorem \\
lemma & theorem \\
corollary & theorem \\
definition & definition \\
proof & proof \\
\bottomrule
\end{tabular}
\end{minipage}
\hfill
\begin{minipage}[t]{.32\linewidth}
\centering
\begin{tabular}{l|l}
\toprule
\textbf{Source} & Textbook: NT \\
\toprule
\textbf{\LaTeX{} env} & \textbf{Type} \\
\midrule
theorem & theorem \\
lemma & theorem \\
corollary & theorem \\
proposition & theorem \\
definition & definition \\
proof & proof \\
\bottomrule
\addlinespace[0.5em]
\end{tabular}
\end{minipage}
\caption{
    Mappings from \LaTeX{} environment names to \textsc{NaturalProofs} data types for each data source.
    As an example, for Stacks, the mapping from \textit{lemma} to \textit{theorem} in row 2 means that an environment enclosed by \texttt{\textbackslash{}begin\{lemma\}} and \texttt{\textbackslash{}end\{lemma\}} is considered a theorem in \textsc{NaturalProofs}.
}
\label{tbl:preprocess}
\end{table}

\begin{figure}[ht]
\begin{minted}[
               linenos=false,
               xleftmargin=2pt,
               tabsize=2,
               fontsize=\scriptsize,
               ]{js}
Content Categories
    ...
    Definitions
        ...
        Definitions by Topic
            ...
            Definitions/Branch of Mathematics
                Definitions/Abstract Algebra
                Definitions/Algebra
                Definitions/Analysis
                ...
                Definitions/Topology
    Proofs
        ...
        Proofs by Topic
            Abstract Algebra
                Additive Functions
                    Examples of Additive Functions
                    Monotone Additive Function is Linear
                Additive Groups
                ...
                Zero Elements
            Algebra
            Analysis
            ...
            Trigonometry
\end{minted}
\caption{
    ProofWiki category graph.
    Nested structure represents sub-categories.
    Some nesting omitted here for simplicity.
}
\label{tbl:category-graph}
\end{figure}
\begin{figure}[ht]
\begin{minipage}{.40\linewidth}
\centering
\includegraphics[width=\columnwidth]{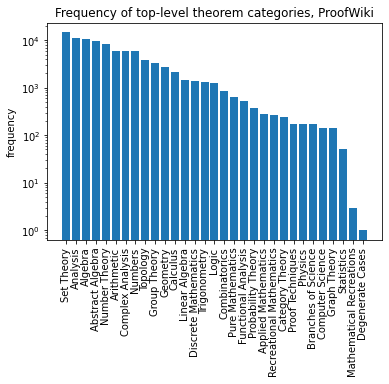}
\caption{
    Frequency of top-level categories, ProofWiki.
}
\label{fig:tlcat-freq}
\end{minipage}
\hfill
\begin{minipage}{.40\linewidth}
\centering
\includegraphics[width=\columnwidth]{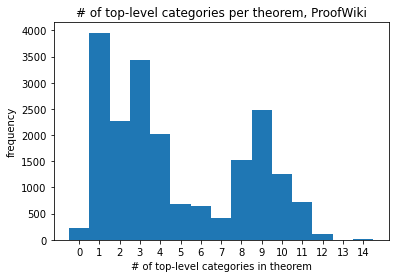}
\caption{
    Number of top-level categories per theorem, ProofWiki.
}
\label{fig:tlcats-per-thm}
\end{minipage}
\end{figure}

\subsection{ProofWiki categories.}
\label{apx:ssec:proofwiki-categories}

For ProofWiki, we also provide category tags for each statement.
ProofWiki contains statements encompassing a broad coverage of mathematical topics (i.e. categories).
In ProofWiki, each category has zero or more sub-categories, and sub-categories have sub-sub-categories, and so on, forming a \textit{category graph}.\footnote{It is not strictly a tree or DAG, because there are several skip connections (e.g. \texttt{Complex Analysis} is both a top-level category and a sub-category under \texttt{Analysis}) and circular dependencies (e.g. \texttt{Metric Spaces} and \texttt{Pseudometric Spaces} are sub-category of each other)}
We recursively scrape the category pages starting from \texttt{Category:Content Categories},\footnote{\url{https://proofwiki.org/wiki/Category:Content_Categories}} and consider categories directly under \texttt{Category:Proofs By Topic} as top-level categories.
\autoref{tbl:category-graph} shows the high-level structure of the ProofWiki category graph.

In the ProofWiki raw data, each statement page is tagged with several categories (the \mintinline{js}{'categories'} field).
In addition, we find the top-level categories (the \mintinline{js}{'toplevel_categories'} field) as well as exhaustive categories (the \mintinline{js}{'recursive_categories'} field) for each theorem by running flood-fill on the category graph.
\autoref{fig:tlcat-freq} and \autoref{fig:tlcats-per-thm} show some statistics of the top-level categories.

\clearpage

\section{Implementation Details and Experimental Setup}
\label{apx:experiments}

\paragraph{Model input format.}
We format each statement ($\xb$ or $\rb$) as,
$\texttt{[CLS]} \text{ title } \texttt{[SEP]} \text{ content } \texttt{[SEP]}$,
and we truncate the statement when the sequence exceeds the model's maximum length.
Each sequence is tokenized using the \texttt{bert-base-cased} tokenizer.

\begin{figure}[h]
\centering
\includegraphics[width=\columnwidth]{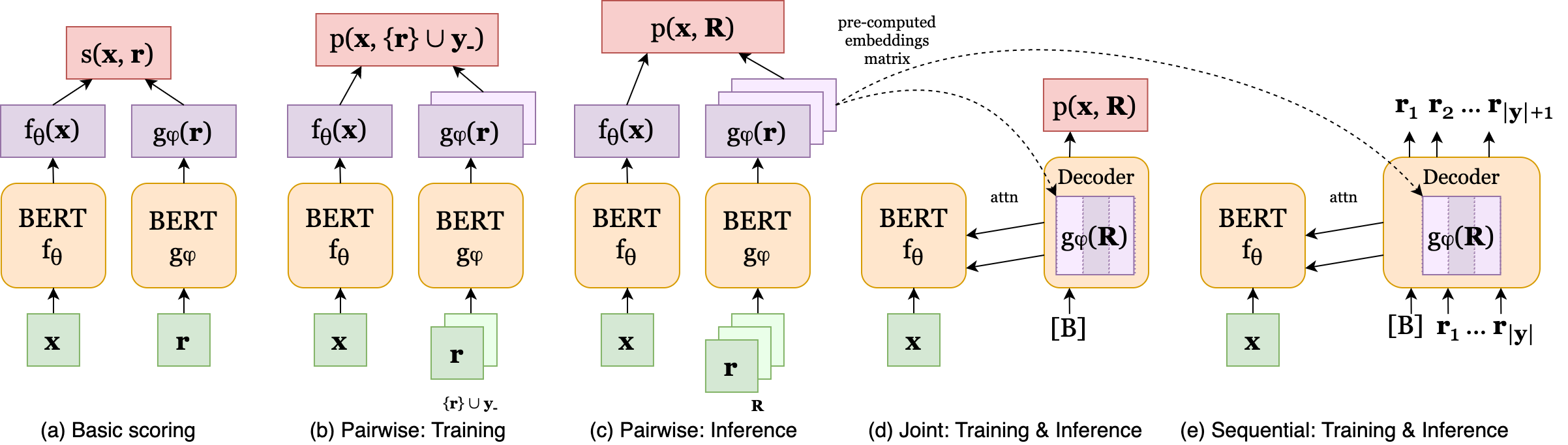}
\caption{
    Model diagrams for the mathematical reference retrieval task.
    (a) The basic pairwise scoring of a theorem ($\xb$) and a reference ($\rb$).
    We use two independently parameterized BERT models ($\mathbf{f}_\theta$ and $\mathbf{g}_\phi$) to encode theorems and references, respectively.
    The theorem embedding $\mathbf{f}_\theta$ and $\mathbf{g}_\phi(\xb)$ and the reference embedding $\mathbf{g}_\phi(\rb)$ are taken to produce a pairwise score $s(\xb, \rb)$.
    (b) The training schema for the pairwise parameterization model.
    A small negative reference set $\yb_-$ is chosen, and the model is trained to maximize the probability that the ground-truth reference is selected.
    (c) The inference schema for the pairwise parameterization model.
    The complete reference set $\mathcal{R}$ is ranked for each theorem.
    (d) The schema for the joint parameterization model.
    The decoder takes the pre-computed embeddings matrix from the pairwise inference step, and does a one-step generation to predict a distribution over the reference set.
    References are ranked based on their probability masses in this distribution.
    (e) The schema for the sequential generation and retrieval model.
    It resembles the joint model, except that its decoder does a multi-step generation to rollout an ordered list of references.
}
\label{fig:arch}
\end{figure}

\subsection{Pairwise model}
\label{apx:ssec:pairwise}
Models are implemented with \texttt{transformers} \citep{wolf2020transformers} and \texttt{pytorch-lightning}\footnote{\url{https://github.com/PyTorchLightning/pytorch-lightning}}.
The theorem encoder $f_{\theta_1}^{\text{thm}}$ is parameterized using the \texttt{bert-base-cased} architecture and initialized with its parameters.
The reference encoder $g_{\theta_2}^{\text{ref}}$ is also parameterized and initialized with (a separate instance of) \texttt{bert-base-cased}.

\paragraph{Training.}
Models are trained for 500,000 steps on one Quadro RTX 8000 GPU.
Each batch contains a maximum of 16,384 ($2^{14}$) tokens.
Validation is done every 5,000 steps. 
The model with the highest mAP computed on the validation set is selected for final evaluation.

\paragraph{Negatives.}
We use \textit{in-batch negatives} as in \citep{karpukhin2020dense}, which computes a score matrix $\mathbf{S}=\mathbf{TR}^\top\in \mathbb{R}^{B\times B}$ on a batch of theorem embeddings $\mathbf{T}\in \mathbb{R}^{B\times d}$ and reference embeddings $\mathbf{R}\in \mathbb{R}^{B\times d}$, then defines the loss as $\sum_{i=1}^B \text{softmax}(\mathbf{S}[i,:])$, which treats elements on the diagonal of $\mathbf{S}$ as positives and off-diagonal elements as negatives.

\paragraph{Evaluation.}
The full set of inputs $\mathbf{x}$ and the full set of references $\mathcal{R}$ are pre-encoded using their respective trained models (i.e. two instances of BERT).
Then the encodings for each possible $\mathbf{x},\mathbf{r}$ pair are used to obtain scalar scores, inducing a ranked list of all $|\mathcal{R}|$ references for each input $\mathbf{x}$.

\subsection{Autoregressive}
\label{apx:ssec:autoregressive}
We implement the autoregressive model as a sequence-to-sequence encoder-decoder model.
Following \citet{rothe2020leveraging}, we parameterize the encoder and decoder using BERT models. 
This allows for initializing with pairwise model components. 
Concretely, we implement the architecture using the \texttt{transformers} \texttt{EncoderDecoderModel} class with \texttt{bert-base-cased} encoder and decoder.

Let $f_{\theta_1}(\mathbf{x})$ denote the encoder and $h_{\theta_2}(\rb_{<t},f_{\theta_1}(\mathbf{x}))$ denote the decoder.
The decoder has an embedding matrix $\mathbf{R}\in \mathbb{R}^{(|\mathcal{R}|+2)\times d}$, where each row represents a reference or special token $\langle bos \rangle$, $\langle eos \rangle$.
At each step $t$, given a theorem and sequence of tokens $(\langle bos \rangle, \rb_1,\ldots,\rb_{t-1})$, the decoder produces a next-token distribution $p_{\theta}(\cdot|\xb,\rb_{<t})=\text{softmax}(\mathbf{R}h_t + \mathbf{b})$, where $h_t\in\mathbb{R}^d$ is the final hidden state obtained from the decoder $h_{\theta_2}(\rb_{<t},f_{\theta_1}(\xb))$, and $\mathbf{b}\in \mathbb{R}^{(|\mathcal{R}|+2)}$ is a bias vector.

The model is trained using cross-entropy loss with the ground-truth $(\xb,\yb)$ pairs, where $\yb=(\langle bos \rangle,\rb_1,\ldots,\rb_{|\yb|},\langle eos \rangle)$ is a reference sequence.

\paragraph{Initialization.}
Let $f^{\text{thm}}_{\tilde{\theta}_1}$ and $g^{\text{ref}}_{\tilde{\theta}_2}$ be the theorem and reference encoder from a trained pairwise model (\S\ref{apx:ssec:pairwise}).
The initialization settings listed in \autoref{tbl:autoregressive-ablation} are as follows. 
$f^{\text{thm}}$ means initializing the encoder $f_{\theta_1}$'s parameters as $\theta_1=\tilde{\theta}_1$, and then updating them during training.
$\mathbf{R}$ means initializing and freezing the decoder's embedding matrix as (omitting the $\langle bos \rangle$ and $\langle eos \rangle$ rows),
\begin{align*}
    \mathbf{R}=\begin{bmatrix}
         \horzbar & g^{\text{ref}}_{\tilde{\theta}_2}(\rb_1) & \horzbar\\
         &\ldots &\\
         \horzbar & g^{\text{ref}}_{\tilde{\theta}_2}(\rb_{|\mathcal{R}|}) & \horzbar\\
\end{bmatrix}.
\end{align*} 

\paragraph{Training.}
Models are trained for 50 epochs on one Quadro RTX 8000 GPU.
Each batch contains a maximum of 16,384 ($2^{14}$) tokens.
Validation is done every 5 epochs. 
The model with the highest mAP computed on the validation set is selected for final evaluation.

\paragraph{Generation evaluation.}
Let $\hat{\yb}\sim \mathcal{F}(p_\theta, \xb)$ denote decoding a sequence $\hat{\yb}=(\rb_1,\ldots,\rb_{|\hat{\yb}|}, \langle eos \rangle)$ given model $p_\theta$ and input $\mathbf{x}$, using decoding algorithm $\mathcal{F}$.
For the reference generation task (\S\ref{ssec:refgen}), we use beam search with beam size 20, based on a preliminary search over beam size \texttt{\{1,10,20,50\}}.
For retrieval evaluation only, we use greedy decoding (beam size 1) with a 1-gram repetition mask since duplicates are not used during retrieval evaluation.
For all decoding algorithms, we use the \texttt{transformers} implementations.

\paragraph{Retrieval evaluation.}
A retrieval model produces a ranked list $\rb^{(1)},\ldots,\rb^{(|\mathcal{R}|)}$ given an input $\xb$.
We evaluate our autoregressive model as a retrieval model by producing a ranked list $\rb^{(1)},\ldots,\rb^{(|\hat{\yb}|)},\ldots,\rb^{(|\mathcal{R}|)}$,
where the first $|\hat{\yb}|$ references come from the model's generated sequence $\hat{\yb}=(\rb^{(1)},\ldots,\rb^{|\hat{\yb}|})$ after removing duplicates, and the remaining references are ordered according to the model's first-step probabilities, $p_{\theta}(\rb_1|\xb,\langle bos \rangle)$.
In preliminary experiments we found the first step's probabilities to perform slightly better than using the last step's probabilities.

\subsection{Joint retrieval}
\label{apx:ssec:joint}
We implement the joint retrieval model as a one-step variant of the autoregressive retrieval model, 
\begin{align}
p_{\theta}(\cdot|\mathbf{x})=\text{softmax}(\mathbf{R}h_t+\mathbf{b}),
\end{align}
where $h_t\in\mathbb{R}^d$ is the final hidden state obtained from $h_{\theta_2}(\langle bos \rangle, f_{\theta_1}(\mathbf{x}))$, and $f_{\theta_1}$, $h_{\theta_2}$ are implemented using the same encoder-decoder architecture as the autoregressive model (\S\ref{apx:ssec:autoregressive}).
This was a design decision made to closely compare the effect of autoregressive vs. joint parameterizations; an alternative implementation could use an encoder-only model.

The model is trained using KL-divergence loss, using per-example reference-distributions $$p_*(\rb|\xb,\yb)=\begin{cases}\frac{1}{|\yb|} & \rb\in \yb \\ 0 & \text{otherwise}\end{cases},$$ where $\yb=\{\rb_1,\ldots,\rb_{|\yb|}\}$ is the ground-truth reference set.

We use the same training settings that were used with the autoregressive model (\S\ref{apx:ssec:autoregressive}).

\subsection{Retrieval Metrics}

For the mathematical reference retrieval task, we evaluate with standard retrieval metrics -- mean average prevision (mAP) and recall@$k$ (R@$k$) -- and a Full@$k$ metric that measures ability to fully recover all true references within the top-$k$ results.
We use $k=10$ and $k=100$ for our evaluation.

\paragraph{mAP.}
Suppose for retrieval example $(\xb, \yb)$ the model ranks all references as $\rb^{(1)},\ldots,\rb^{(|\mathcal{R}|)}$.
The average precision is computed as
\begin{align*}
\mathrm{AP} &=
\sum_{j=1}^{|\mathcal{R}|}{\mathbb{I}[\rb^{(j)} \in \yb] \frac{\sum_{k=1}^{j}{\mathbb{I}[\rb^{(k)} \in \yb]}}{j}}.
\end{align*}
mAP is the mean of AP across all retrieval examples.

\paragraph{R@$k$.}
For each retrieval example, the recall@$k$ is
\begin{align*}
\mathrm{R}@k &=
\frac{\sum_{j=1}^{k}{\mathbb{I}[\rb^{(j)} \in \yb]}}{|\yb|}.
\end{align*}
We aggregate recall@$k$ by micro-averaging across retrieval examples.

\paragraph{Full@$k$.}
For each retrieval example, the fully-recovering indicator is formally defined as
\begin{align*}
\mathrm{Full}@k &=
\prod_{\rb \in \yb}{\mathbb{I} \big[ \rb \in \{ \rb^{(j)} \mid 1 \le j \le k \} \big]}.
\end{align*}
The overall Full@$k$ metric is thus the mean of this fully-recovering indicator across all retrieval examples.

\section{Additional Results}

\begin{table*}[h]
\setlength{\tabcolsep}{4pt}
\begin{center}
\resizebox{\linewidth}{!}{
\begin{tabular}{r | rrrrr | rrrrr}
\toprule
 & \multicolumn{5}{c}{\textbf{ProofWiki}} & \multicolumn{5}{c}{\textbf{Stacks}} \\
\toprule
& \textbf{mAP} & \textbf{R@10} & \textbf{R@100} & \textbf{Full@10} & \textbf{Full@100} & \textbf{mAP} &  \textbf{R@10} & \textbf{R@100} & \textbf{Full@10} & \textbf{Full@100} \\
\toprule
\textbf{Random}     & 0.04 & 0.00 & 0.33 & 0.00 & 0.00      & 0.08 & 0.10 & 0.43 & 0.00 & 0.13  \\
\textbf{Frequency}  & 3.54 & 5.99 & 24.44 & 0.88 & 2.28     & 1.03 & 1.86 & 10.86 & 0.13 & 2.19  \\
\textbf{TF-IDF}     & 6.33 & 10.31 & 21.82 & 4.74 & 8.69    & 13.45 & 24.95 & 48.24 & 19.61 & 36.77 \\
\midrule
\textbf{BERT-pair (P+S)} & 13.84 & 19.31 & 56.99 & 8.60 & 31.96  & 17.29 & 33.29 & 74.14 & 23.61 & 63.23 \\
+\textbf{joint} & 33.85 & 37.15 & 72.25 & 17.12 & 48.46 & 25.12 & 36.00 & 74.24 & 27.35 & 64.13\\
\textbf{BERT-pair} & 16.99 & 22.91 & 62.03 & 9.22 & 36.96  & 21.21 & 38.00 & 75.67 & 28.77 & \textbf{66.19} \\
+\textbf{joint}& \textbf{37.51} & \textbf{41.39} & \textbf{75.92} & \textbf{20.54} & \textbf{50.75} & \textbf{26.55} & \textbf{39.81} & \textbf{75.71} & \textbf{30.58} & 66.06\\
\bottomrule
\end{tabular}}
\end{center}
\caption{
    \textit{In-domain} performance on the mathematical reference retrieval task (validation set).
    \textbf{BERT} is finetuned on the part of dataset with the same source as the evaluation set, whereas \textbf{BERT (P+S)} is finetuned on the combined dataset from ProofWiki and Stacks sources.
    Recall is micro-averaged.
}
\label{tbl:retrieval-main-valid}
\end{table*}

\begin{table}[h]
\setlength{\tabcolsep}{4pt}
\begin{center}
\begin{tabular}{r | llll | llll}
\toprule
 & \multicolumn{4}{c}{\textbf{ProofWiki}} & \multicolumn{4}{c}{\textbf{Stacks}} \\
\toprule
& \textbf{All} & \textbf{Theorems} & \textbf{Definitions} & \textbf{Others} & \textbf{All} & \textbf{Theorems} & \textbf{Definitions} & \textbf{Others} \\
\midrule
\textbf{Frequency} & 3.54 & 7.25 & 5.02 & 1.49   & 1.03 & 1.14 & 0.33 & 0.48 \\
\textbf{TF-IDF}    & 6.33 & 10.07 & 2.33 & 2.19  & 13.45 & 12.11 & 15.51 & 13.94 \\
\textbf{BERT}      & \textbf{16.99} & \textbf{14.71} & \textbf{13.39} & \textbf{11.06}   & \textbf{21.21} & \textbf{19.31} & \textbf{24.39} & \textbf{17.10} \\
\bottomrule
\end{tabular}
\end{center}
\caption{
    Retrieval performance (mAP) by reference type (validation set).
}
\label{tbl:retrieval-type}
\end{table}
\paragraph{Performance by reference type.}
In \autoref{tbl:retrieval-type} we break down the in-domain retrieval performance by reference type.
BERT shows a consistent improvement over TF-IDF on all types of references.
On ProofWiki, TF-IDF does much worse on definitions and other types than on theorems, whereas BERT gives a more balanced performance on different types of references.

\section{Supplementary Materials}

\paragraph{Dataset documentation and intended uses.}
We use the Dataset Nutrition Labels framework \cite{holland2018dataset} for dataset documentation.
For the Statistics module, please refer to \autoref{tbl:dataset-stats}, \autoref{fig:tlcat-freq} and \autoref{fig:tlcats-per-thm}.

\begin{table}[h]
\begin{minipage}[t]{.52\linewidth}
\setlength{\tabcolsep}{3pt}
\setlength{\fboxrule}{1.5pt}
\resizebox{\columnwidth}{!}{
\fbox{
\begin{tabular}{lr}
\multicolumn{2}{l}{\textbf{Metadata}} \\
\toprule
\textbf{Filename} & \texttt{proofwiki.json} \\
                  & \texttt{stacks.json} \\
                  & \texttt{ra-trench.json} \\
                  & \texttt{nt-stein.json} \\
\midrule
\textbf{Format}   & json \\
\midrule
\textbf{Url}      & \url{https://doi.org/10.5281/zenodo.4632538} \\
\midrule
\textbf{Domain}   & natural language processing \\
\midrule
\textbf{Keywords} & mathematics, theorems, proofs, language \\
\midrule
\textbf{Type}     & \\
\midrule
\textbf{Rows}     & 80,795 \\
\midrule
\textbf{Columns}  & 9 \\
\midrule
\textbf{Missing}  & none \\
\midrule
\textbf{License}  & CC BY-SA 4.0 (\texttt{proofwiki.json}) \\
                  & CC BY-NC-SA 4.0 (\texttt{ra-trench.json})  \\
                  & GFDL 1.2 (\texttt{stacks.json})  \\
                  & MIT License (\texttt{ra-stein script})  \\
\midrule
\textbf{Released} & June 2021 \\
\midrule
\textbf{Range}    & N/A \\
\midrule
\textbf{Description} & This dataset is a collection of mathematical \\
                     & statements and proofs in natural language. \\
                     & It collects data from multiple sources, \\
                     & encompassing broad-coverage of all math \\
                     & topics, deep-dive with a selected topic, and \\
                     & low-resource scenarios. The dataset provides \\
                     & theorems, proof(s) to each theorem when \\
                     & applicable, and in-proof references to other \\
                     & mathematical statements. \\
\bottomrule
\end{tabular}
}
}
\end{minipage}
\hfill
\begin{minipage}[t]{.48\linewidth}
\setlength{\tabcolsep}{3pt}
\setlength{\fboxrule}{1.5pt}
\resizebox{\columnwidth}{!}{
\fbox{
\begin{tabular}{lr}
\multicolumn{2}{l}{\textbf{Provenance}} \\
\toprule
\textbf{Source} \\
                \multicolumn{2}{l}{ProofWiki} \\
                & (\url{https://proofwiki.org/}) \\
                \multicolumn{2}{l}{Stacks} \\
                & (\url{https://stacks.math.columbia.edu/}) \\
                \multicolumn{2}{l}{Textbook: Real Analysis} \\
                & (\url{https://digitalcommons.trinity.edu/mono/7/}) \\
                \multicolumn{2}{l}{Textbook: Number Theory} \\
                & (\url{https://wstein.org/ent/}) \\
\midrule
\textbf{Author} \\
Name & Sean Welleck et al. \\
Email & \url{wellecks@uw.edu} \\
\bottomrule
\end{tabular}
}
}
\resizebox{\columnwidth}{!}{
\fbox{
\begin{tabular}{lr}
\multicolumn{2}{l}{\textbf{Variables}} \\
\toprule
\textbf{id} & A unique ID for this statement. \\
\midrule
\textbf{type} & The type of this statement; \\
              & either \textit{theorem}, \textit{definition}, or \textit{other}. \\
\midrule
\textbf{label} & A string description of this statement. \\
\midrule
\textbf{categories} & A list of topics that this statement \\
                    & pertains. For ProofWiki data only. \\
\midrule
\textbf{title} & A descriptive title of this statement. \\
\midrule
\textbf{contents} & The content of this statament or \\
                  & proof, written in \LaTeX{}. \\
\midrule
\textbf{refs} & A list of labels of statements that this \\
              & statement or proof refers to in its content. \\
\midrule
\textbf{ref\_ids} & IDs for items in \texttt{refs}. \\
\midrule
\textbf{proofs} & A list of proofs for this theorem. \\
                & May be empty. \\
\bottomrule
\addlinespace[0.5em]
\end{tabular}
}
}
\end{minipage}
\caption{Dataset Nutrition Labels for \textsc{NaturalProofs}.}
\label{tbl:ds-doc}
\end{table}

The \textsc{NaturalProofs} dataset is intended to be used by researchers to build or evaluate machines on predicting references in proofs, generating proofs to mathematical theorems, or other related tasks.
It should not be regarded as source of truth for defining particular mathematical concepts, proving particular mathematical theorems, or the existence of such proof(s).
In that case the user is advised to consult authoritative mathematical resources.

\paragraph{Dataset URL.}
The \textsc{NaturalProofs} dataset is hosted at \url{https://doi.org/10.5281/zenodo.4632538}.
Additional instructions and resources are provided in the Github repo \url{https://github.com/wellecks/naturalproofs}.

\paragraph{Author statement and license.}
We bear all responsibility in case of violation of rights.
We confirm that the data sources we use are licensed to permit redistribution with modification for non-commercial purposes.

\paragraph{Hosting, licensing, and maintenance plan.}
The dataset is hosted and maintained through Zenodo~\citep{zenodo},\footnote{\url{https://zenodo.org/}} and the code is hosted by GitHub.
The code is released under the MIT license.
The dataset is released under per-file licenses: CC BY-SA 4.0 (\texttt{proofwiki.json}), CC BY-NC-SA 4.0 (\texttt{ra-trench.json}), GFDL 1.2 (\texttt{stacks.json}), MIT License (\texttt{ra-stein script}).
Zenodo meta-data is openly available under the CC0 license, and all open content is openly accessible through open APIs.\footnote{\url{https://about.zenodo.org/}}

\paragraph{Links to access the dataset and its metadata.}
The \textsc{NaturalProofs} dataset is hosted at \url{https://doi.org/10.5281/zenodo.4632538}.
Additional instructions and resources are provided in the Github repo \url{https://github.com/wellecks/naturalproofs}.

\paragraph{Data format.}
We store the dataset as JSON files.
The dataset can be read using common JSON libraries (e.g. the built-in \texttt{json} module in Python) and following the dataset schema in \autoref{fig:schema}.

\paragraph{Long-term preservation.}
We ensure this by uploading the dataset to the Zenodo dataset repository.

\paragraph{Explicit license.}
The code is released under the MIT license.
The dataset is released under per-file licenses: CC BY-SA 4.0 (\texttt{proofwiki.json}), CC BY-NC-SA 4.0 (\texttt{ra-trench.json}), GFDL 1.2 (\texttt{stacks.json}), MIT License (\texttt{ra-stein script}).
Zenodo meta-data is openly available under the CC0 license, and all open content is openly accessible through open APIs.

\paragraph{Structured metadata.}
We release the metadata along with the dataset on Zenodo.

\paragraph{Persistent dereferenceable identifier.}
\url{https://doi.org/10.5281/zenodo.4632538}.

\paragraph{Reproducibility.}
We ensure this by releasing our code on GitHub, which includes instructions to reproduce the evaluation numbers in the paper.

\end{document}